\documentclass[journal]{IEEEtran}

\usepackage{cite}
\usepackage{array}
\usepackage{color}
\usepackage{float}
\usepackage[cmex10]{amsmath}
\usepackage{nomencl}						
\usepackage[normalem]{ulem}			
\usepackage{diagbox}						
\usepackage{slashbox}						
\usepackage{colortbl}						
\usepackage{multirow}						
\usepackage{tabularx}
\usepackage{placeins}
\usepackage{algorithm}
\usepackage[noend]{algpseudocode}
\usepackage{siunitx}
\usepackage{gensymb}
\usepackage{arydshln}
\usepackage{stfloats}
\usepackage{amssymb}
\usepackage{url}
\usepackage{xcolor}
\usepackage{nomencl}

\ifCLASSINFOpdf
  \usepackage[pdftex]{graphicx}
	\graphicspath{{./figure/}}
  \DeclareGraphicsExtensions{.pdf,.jpeg,.png}
\else
  \usepackage[dvips]{graphicx}
  \graphicspath{{./figure/}}
  \DeclareGraphicsExtensions{.eps}
\fi

\ifCLASSOPTIONcompsoc
  \usepackage[caption=false,font=normalsize,labelfont=sf,textfont=sf]{subfig}
\else
  \usepackage[caption=false,font=footnotesize]{subfig}
\fi

\newcommand{\figref}[1]{\textcolor{black}{Fig.~\ref{#1}}}

\newcommand{\sectionref}[1]{\textcolor{black}{Section~\ref{#1}}}
\newcommand{\appendixref}[1]{\textcolor{black}{Appendix~\ref{#1}}}

\begin{document}
\bstctlcite{IEEEexample:BSTcontrol}
\setcounter{page}{1}
\onecolumn
{\Large This work has been submitted to the IEEE for possible publication. Copyright may be transferred without notice, after which this version may no longer be accessible.}
\clearpage

\twocolumn
\title{Eigenvalue Patterns and Participation Analysis of Symmetric Renewable Energy Power Systems}
\author{Yao Qin, \IEEEmembership{Graduate Student Member, IEEE}, Yitong Li, \IEEEmembership{Member, IEEE}, Wei Wang, Shaoze Zhou, Zheng Wei, \IEEEmembership{Member, IEEE}, Jinjun Liu, \IEEEmembership{Fellow, IEEE}
\thanks{Corresponding Author: Yitong Li.}
\thanks{Yao Qin, Yitong Li, Wei Wang, Jinjun Liu are with the State Key Laboratory of Electrical Insulation and Power Equipment, School of Electrical Engineering, Xi'an Jiaotong University, Xi'an 710049, China. E-mail: yaoqin@stu.xjtu.edu.cn; yitongli@xjtu.edu.cn; wangwei@sgepri.sgcc.com.cn; jjliu@mail.xjtu.edu.cn.}
\thanks{Wei Wang, Shaoze Zhou, Zheng Wei are with the NARI Technology Co., Ltd., China. E-mail: wangwei@sgepri.sgcc.com.cn; zhoushaoze@sgepri.sgcc.com.cn; weizheng2@sgepri.sgcc.com.cn.}
\thanks{This work was supported by the National Natural Science Foundation of China under Grant 52307222.}
}

\ifCLASSOPTIONpeerreview
	\maketitle 
\else
	\maketitle
\fi


\begin{abstract}
State-space analysis is widely employed for examining power system dynamics but faces challenges in large-scale power systems integrated with numerous inverter-based resources (IBRs), where the significant increase of system states complicates modal analysis. Notably, renewable energy power systems often consist of multiple homogeneous generation units. This uniformity, termed \textit{symmetry} in this paper, can facilitate the system stability analysis. Eigenvalue patterns and participation factors in three types of symmetric renewable energy power systems are investigated, including ideally-, quasi-, and group-symmetric systems. An ideally-symmetric (quasi-symmetric) system comprises a group of identical (similar) subsystems connected to an external grid. A system containing multiple such groups is termed group-symmetric. In these symmetric systems, two types of modes are defined to characterize different interactions: \textit{inner-group modes}, which describe the interactions among subsystems within a single group, and \textit{group-grid modes}, which describe the interactions between the groups and the external grid. A new concept termed \textit{group participation factor} is also proposed to extend the use of conventional participation factors for repeated and close modes. In addition, the invariance properties of the inner-group modes and group-grid modes are discussed. The findings provide insights for stability analysis and targeted optimization in power systems. Theoretical advances are validated through numerical results and electromagnetic transient (EMT) simulations on example power systems of varied types and scales.
\end{abstract}


\begin{IEEEkeywords}
Symmetry, participation factor, eigenvalue pattern, modal analysis, small-signal stability, renewable energy power system.
\end{IEEEkeywords}

\setlength{\nomlabelwidth}{1.5cm}
\setlength{\nomitemsep}{0.3em}
\makenomenclature
\nomenclature[01]{$P$}{Transformation matrix}
\nomenclature[02]{$\text{eig}(\cdot)$}{Eigenvalue operator}
\nomenclature[03]{$A,a_{kk}$}{State matrix and its $k$-th diagonal element}
\nomenclature[04]{$\Psi,\Phi$}{Left and right eigenvector matrices}
\nomenclature[05]{$pf$}{Participation factor}
\nomenclature[06]{$PF$}{Participation factor matrix}
\nomenclature[07]{$gpf$}{Group participation factor}
\nomenclature[08]{$GPF$}{Group participation factor matrix}
\nomenclature[09]{$M$}{Number of subsystems within a group}
\nomenclature[10]{$N$}{Number of groups}
\nomenclature[11]{$n_g$}{Number of repeated modes or close modes}
\nomenclature[12]{$c$}{Model modification factor}
\nomenclature[13]{$RC$}{Relative change of a eigenvalue}
\nomenclature[14]{SG}{Synchronous generator}
\nomenclature[15]{IBR}{Inverter-based resource}
\nomenclature[16]{WF}{Wind farm}
\nomenclature[17]{PV}{Photovoltaic}
\nomenclature[18]{BESS}{Battery energy storage system}
\nomenclature[19]{EMT}{Electromagnetic transient}
\nomenclature[20]{PCC}{Point of common coupling}
\nomenclature[21]{GFM}{Grid-forming}
\nomenclature[22]{GFL}{Grid-following}
\nomenclature[23]{PLL}{Phase-locked loop}
\nomenclature[24]{LPF}{Low-pass filter}
\nomenclature[25]{GFM-ESS}{Grid-forming energy storage system}
\nomenclature[26]{HFO}{High frequency oscillation}
\nomenclature[27]{MFO}{Medium frequency oscillation}
\nomenclature[28]{LFO}{Low frequency oscillation}

\ifCLASSOPTIONonecolumn
	\printnomenclature[2cm]
\else
	\printnomenclature
\fi


\section{Introduction} \label{Section:Introduction}

\IEEEPARstart{I}{n} recent years, power systems are undergoing significant and complex changes. These changes are driven by growing environmental concerns, the depletion of conventional energy resources, rapid technological advancements, and increasing electricity demand\cite{blaabjerg2006overview}. One of the most critical shifts is the transition from synchronous generators (SGs) towards inverter-based resources (IBRs), such as wind generation, photovoltaic (PV) panels, battery energy storage systems (BESSs), etc. This trend has introduced more stability threats across a wide range of frequencies\cite{wang2019harmonic}.

A general approach for conducting stability analysis is based on the state-space model\cite{kundur1994power,pogaku2007modeling}. To be more specific, system stability is predicted by examining the eigenvalues (or modes) of the state matrix of the entire system. The participation factor, which quantifies the contribution of each state variable to a particular eigenvalue, is derived from the corresponding eigenvectors. Since state variables represent the dynamics of specific physical components within the system, participation factor analysis allows us to trace the root causes of instability back to these physical components and implement targeted mitigation\cite{kundur1994power,li2024descriptor}. State-space analysis has not only been successfully applied in conventional SG-dominated power systems, but is also increasingly used in IBR-based grids. These include inverter-based microgrids\cite{pogaku2007modeling}, parallel-connected inverters\cite{coelho2002small-signal}, and wind farms\cite{kunjumuhammed2017stability}.

In recent years, advanced state-space modeling techniques have emerged, such as descriptor\cite{li2024descriptor}, continuous-time stochastic\cite{wan2025continuous}, discrete-time construction\cite{han2022discrete}, linear time-periodic\cite{zhu2021small}, and nonlinear modular modeling\cite{cecati2022nonlinear}. Concurrently, eigenvalue and participation analysis have also seen substantial progress. New techniques enable the estimation of participation factors directly from measurement data\cite{xia2025estimation}, and the concept has been extended to impedance-based models, offering a grey-box stability assessment approach\cite{zhu2022participation}. Furthermore, the theoretical foundations have been refined, for example, by introducing participation factors for algebraic variables\cite{tzounas2020modal} and proposing a new definition for state-in-mode participation factors\cite{iskakov2021definition}. However, with high penetration of IBRs, there would be a dramatic increase of the number of state variables for the whole system. Consequently, the resulting state-space system model contains a large number of eigenvalues and participation factors, rendering modal recognition and participation interpretation more complicated\cite{kunjumuhammed2017adequacy,Amin2017small-signal}.

To address the aforementioned problem, it is noticed that renewable energy power systems often consist of dozens or even hundreds of nearly identical renewable energy generation units. This uniformity introduces certain characteristics which are helpful for state-space modeling and stability analysis. In PV power plants, for example, \cite{agorreta2011modeling} and \cite{ackermann2016stability} have shown that identical $N$-parallel inverters connected to a grid impedance can be modeled as a single inverter connected to a grid impedance magnified $N$ times. Both models capture the same inverter-grid interaction characteristics. A similar modeling approach has been applied to wind farms when analyzing wind-farm-grid interactions. As shown in \cite{du2019dynamic} and \cite{shao2021an}, when studying inside-wind-farm dynamics, identical $N$-paralleled units can be represented as a single unit connected to an infinite busbar. These equivalent models, though simplified, are effective tools for small-signal stability analysis. Besides modeling, it is revealed in \cite{wu2016a} that homogeneous wind farms tend to exhibit multiple sub-synchronous oscillation modes that are closely located. Building on this phenomenon, \cite{shao2023participation} further demonstrated that these close modes would result in highly sensitive participation factors and would even probably make the participation analysis failed. For a multi-infeed power electronic system with homogeneous inverters, \cite{dong2019small} proposes a generalized short circuit ratio to assess grid strength and simplify stability analysis. While prior studies provide valuable insights, the eigenvalue patterns and participation distribution in renewable energy power systems are not explicitly derived or analyzed. Furthermore, most existing researches focus on a specific scene, without discussing the generality of the underlying physical principles.
\begin{figure}[t]
\centering
\includegraphics[scale=0.45]{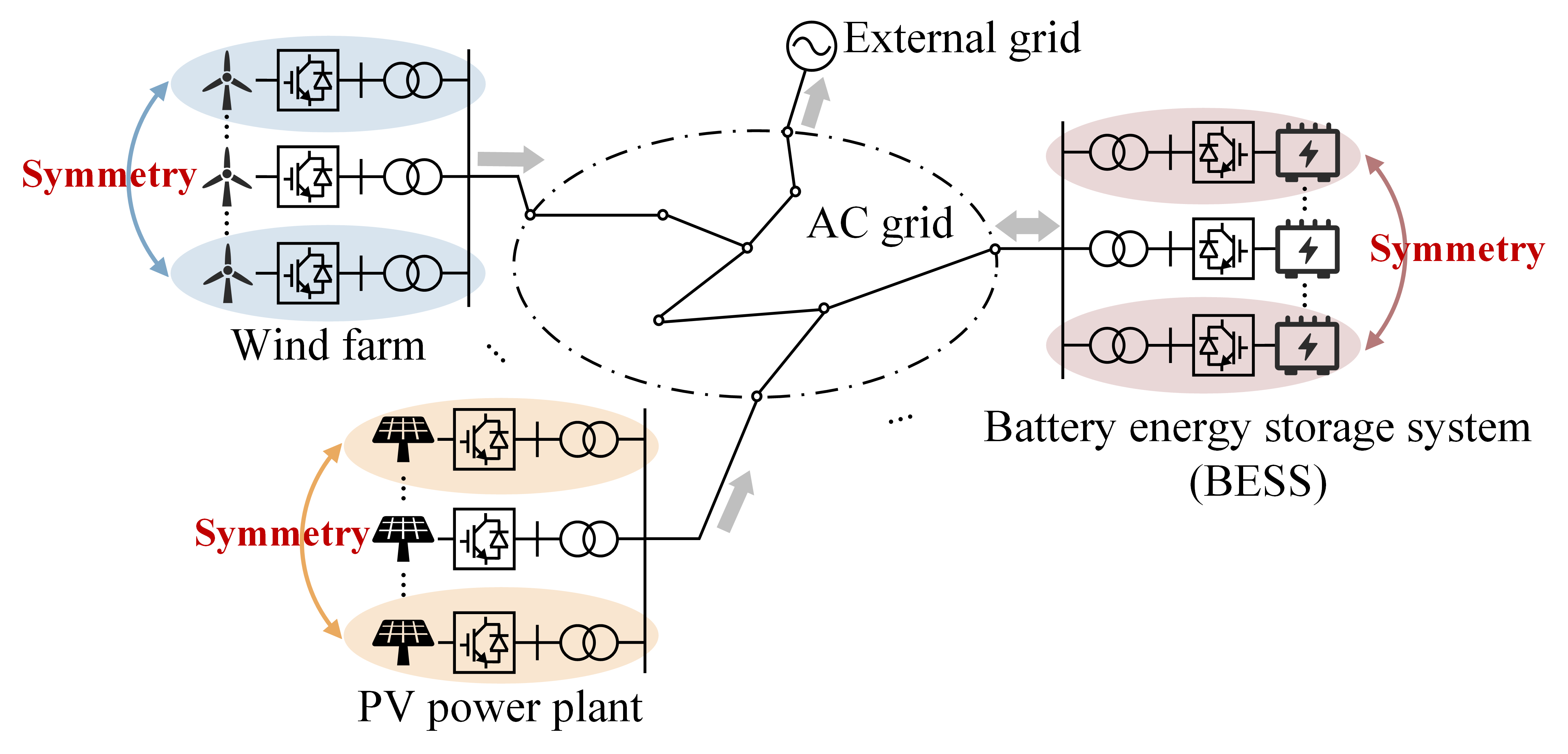}
\caption{Illustration of symmetry in renewable energy power systems.}
\label{Fig:Introduction}
\end{figure}

In this paper, the uniformity in renewable energy power systems is renamed as \textit{symmetry}. The concept of symmetry was originally proposed in physics. An object is considered symmetric under a given operation if it remains unchanged after that operation, and this operation is termed a symmetry operation\cite{griffiths2008introduction,brading2003symmtries}. Depending on different operations, there are different symmetries. If the operation is continuous, such as time translation, spatial translation, or spatial rotation, the resulting symmetry is continuous symmetry. According to Noether's theorem, each continuous symmetry corresponds to a conservation law and yields a conserved quantity\cite{banados2016a}. In contrast, discrete symmetry arises from discrete operations like parity or time reversal, and is prominent in fields such as particle physics and quantum field theory. Although the discrete symmetry may also lead to a conserved quantity in some cases, such quantity is distinct from Noether-conserved quantity and is not guaranteed to exist. Through the framework of symmetry and conservation, the intrinsic principles underlying complex phenomena can be better understood.

This paper extends the aforementioned concept of symmetry to electrical power systems. As illustrated in \figref{Fig:Introduction}, a system is considered symmetric if interchanging any two generation units within a wind farm does not alter its performance, similarly for PV power plants, or BESSs. Here, each generation unit (e.g., for a wind farm) comprises a wind turbine generator, converter, and step-up transformer as an integrated whole. System performance refers to both the steady-state operating points and dynamic responses. This paper focuses specifically on the small-signal stability aspect of system performance. But obviously in practice, the system symmetry is not always ideally valid. On this basis, symmetric systems are further classified as ideally-, quasi-, and group-symmetric systems in this paper. The eigenvalue patterns of the three types of symmetric systems are studied, leading to the identification of \textit{inner-group modes} and \textit{group-grid modes}, which respectively represent the inner-group dynamics and group-grid interactions. The corresponding participation factors of these two types of modes are also examined. In this context, a new concept termed \textit{group participation factor} is proposed to provide a more refined evaluation of component contributions to each mode. Furthermore, by drawing on the idea of symmetry and conservation, it is demonstrated that the two mode types remain almost unchanged under certain circumstances. The group participation factors and invariance properties of modes offer a theoretical basis for mitigating unstable modes in a more targeted and effective manner.

The main contributions of this paper can be summarized as follows:

\begin{itemize}
    \item [1)]
    By extending the concept of symmetry from physics, this paper introduces a novel framework of symmetry in renewable energy power systems, classifying them as ideally-, quasi-, and group-symmetric. This framework provides a new perspective for analyzing and enhancing small-signal stability of power systems.
    \item [2)]
    Using similarity transformation, the eigenvalue patterns of these symmetric systems are defined. Specifically, inner-group modes are always identified as repeated or close modes, while group-grid modes are always single and distinct modes. It is also proved that these modes would remain nearly invariant under certain circumstances.
    \item [3)]
    Conventional participation factors are ineffective for analyzing repeated and close modes. To address this limitation, a group participation factor is proposed, which offers an easy but effective way to quantify the component contributions to these modes.
\end{itemize}

In addition, it is important to distinguish the inner-group modes and group-grid modes from established concepts in the literature: (i) inside-wind-farm/wind-farm-grid\cite{shao2023participation}, (ii) internal/external\cite{du2019dynamic}, and (iii) local/inter-area modes\cite{afaq2024nonlinear}. First, modes (i) are specific to wind farms and the station level, whereas inner-group and group-grid modes are applicable to a wide range of scenarios and focus on the group level. The two concepts operate at a different scale, as a single wind farm may contain multiple groups. Second, the boundary between ``internal'' and ``external'' in modes (ii) is relative, whereas a ``group'' is well-defined with clear physical significance. Finally, modes (iii) typically describe low frequency oscillations in SG-dominated systems, whereas inner-group and group-grid modes can characterize wide-band oscillations in IBR-based grids. Therefore, while inspired by established ideas, the proposed concepts offer enhanced clarity, broader applicability, and greater relevance to renewable energy power systems.

The remainder of this paper is organized as follows. Section II investigates the eigenvalue patterns and participation factors of ideally-symmetric renewable energy power systems. It begins with a simple example to facilitate the understanding of viewpoints and then gives a more general derivation. In Section III, modal analysis is conducted on more realistic scenarios: quasi- and group-symmetric renewable energy power systems. Section IV extends the symmetry analysis by proposing group participation factors and modal invariance properties, and illustrates how to apply the proposed theories. Section V presents case studies with numerical results and electromagnetic transient (EMT) simulations to validate the theoretical findings. Finally, Section VI concludes this paper.


\section{Ideally-Symmetric Systems} \label{Section:Ideally-Symmetric Systems}
In this section, a special symmetric system, referred to as an ideally-symmetric system, is investigated. This system operates under idealized conditions that are beneficial for theoretical analysis. To facilitate understanding, we begin with a simple example, followed by a more general derivation.
\subsection{Simple Example} \label{Section: Simple Example}
\begin{figure*}[!ht]
\centering
\includegraphics[scale=0.87]{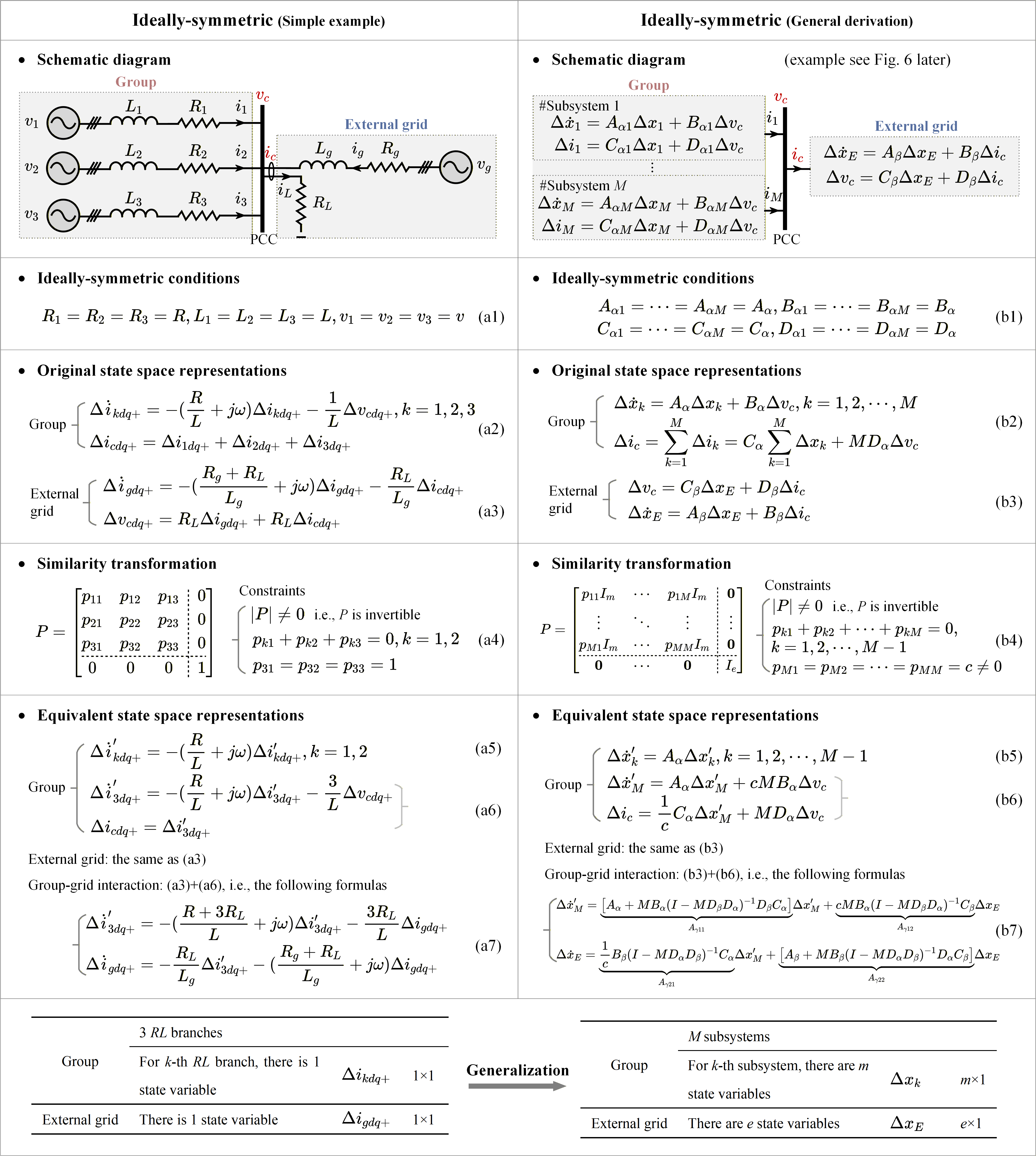}
\caption{State space modeling of ideally-symmetric systems. (Notes: In this paper, the prefix ``$\Delta$'' denotes small-signal perturbations, the superscript ``$^\prime$'' refers to variables that have been transformed by the transformation matrix $P$, and $I$ represents the identity matrix, where the subscript specifies its dimension.)}
\label{Fig:IdeallySymmetric_StateSpace}
\end{figure*}

As depicted in the left column of \figref{Fig:IdeallySymmetric_StateSpace}, a group of three $RL$ branches is in parallel to the external grid. Assuming that these $RL$ branches are identical and connected to the same voltage source, as shown in (a1), the entire grid-connected system can be regarded as an ideally-symmetric system. The original state space representations are provided in (a2) and (a3). The synchronous $dq$ reference frame in complex vector form is used\cite{harnefors2007modeling}. The dynamics associated with the system frequency $\omega$ ($\Delta \omega$) and the backward components ($dq-$) are not considered, so that each $RL$ branch is characterized by a single state variable $\Delta i_{dq+}$.

To facilitate a more intuitive modal analysis, a transformation matrix $P$ is performed on the state variables. This transformation is mathematically equivalent to a similarity transformation on the state matrix and, therefore, does not alter its eigenvalues\cite{similarity}. The elements of $P$, along with their constraints, are defined in (a4). Notably, the second row of (a4) is employed to eliminate $\Delta v_{cdq+}$ in (a2), while the last row of (a4) establishes a direct relationship between the transformed state variable $\Delta i^{\prime}_{3dq+}$ and the output variable $\Delta i_{cdq+}$. Following the transformation, (a2) is reformulated as (a5) and (a6), whereas (a3) remains unchanged. The method for the equivalent transformation of the state variables and the state matrix refers to \cite{du2019dynamic,shao2021an}, with modifications made in this work. From the derived state space models, the eigenvalue and participation factor analyses are conducted next.

1) \emph{Eigenvalue Patterns}: The eigenvalues of (a5) are computed as
\begin{equation}
    \lambda_{1}=-(\frac{R}{L}+j\omega),\lambda_{2}=-(\frac{R}{L}+j\omega).
    \label{RLInnerMode}
\end{equation}
Note that these two eigenvalues are identical and independent of the dynamics of the external grid, as (a5) does not include the input variable $\Delta v_{cdq+}$. Consequently, these eigenvalues are defined as inner-group modes. Both of them also characterize the modes of the entire grid-connected system.

\begin{figure*}[!hb]
    \centering
    \hrulefill
    \vspace*{1pt}
    \setcounter{equation}{3}
    \begin{equation}
        \label{RLStateMatrix}
        \begin{bmatrix}
        \Delta \dot{i}_{1dq+}  \\
        \Delta \dot{i}_{2dq+}  \\
        \Delta \dot{i}_{3dq+}  \\
        \Delta \dot{i}_{gdq+}
        \end{bmatrix}
        =
        \underbrace{\begin{bmatrix}
            o & r & r & r \\
            r & o & r & r \\
            r & r & o & r \\
            s & s & s & t
        \end{bmatrix}}_{A}
        \begin{bmatrix}
        \Delta i_{1dq+}  \\
        \Delta i_{2dq+}  \\
        \Delta i_{3dq+}  \\
        \Delta i_{gdq+}
        \end{bmatrix}, \text{where}
        \left\{
        \begin{aligned}
             o&=-(\frac{R}{L}+\frac{R_L}{L}+j\omega), r=-\frac{R_L}{L} \\
             s&=-\frac{R_L}{L_g}, t=-(\frac{R_g}{L_g}+\frac{R_L}{L_g}+j\omega)
        \end{aligned}
        \right.
    \end{equation}
\end{figure*}

In contrast, the state space representation in (a6) incorporates interactions with the external grid through the input variable $\Delta v_{cdq+}$. As a result, the eigenvalues derived solely from (a6) do not represent the true modes of the complete grid-connected system. However, by combining (a6) with the external grid model given in (a3), the remaining state space model of the system can be constructed, denoted as (a7). From (a7), the eigenvalues are calculated, denoted as $\lambda_3$ and $\lambda_g$:
\setcounter{equation}{1}
\begin{equation}
    \begin{bmatrix}
        \lambda_{3}& \\
         & \lambda_{g}
    \end{bmatrix}
    =\text{eig}\left (
    \begin{bmatrix}
        -\frac{R+3R_L}{L}-j\omega & -\frac{3R_L}{L} \\
        -\frac{R_L}{L_g} & -\frac{R_g+R_L}{L_g}-j\omega
    \end{bmatrix}\right )
    \label{RLOuterMode}
\end{equation}
where $\text{eig}(\cdot)$ denotes the eigenvalue operator. The eigenvalues of this model are referred to as group-grid modes, as they are influenced by both the $RL$ group and the external grid. Further analysis is conducted by rewriting the first equation of (a7) in another form as
\begin{equation}
    \Delta \dot{i}^{\prime}_{3dq+}=-(\frac{R/3+R_L}{L/3}+j\omega)\Delta i^{\prime}_{3dq+}-\frac{R_L}{L/3}\Delta i_{gdq+}.
    \label{RLOuterInvariantMode}
\end{equation}
In \eqref{RLOuterInvariantMode}, $R/3$ and $L/3$ are the equivalent resistance and inductance of the parallel-connected $RL$ group, respectively. This indicates that the group-grid modes characterize the interactions between the group's aggregated dynamic and the external grid's dynamic. In other words, the group-grid modes are more strongly associated with the equivalent behavior of the group as a whole, rather than with the internal dynamics of its individual components.

\begin{figure*}[!hb]
    \centering
    \hrulefill
    \vspace*{1pt}
    \setcounter{equation}{7}
    \begin{equation}
        \label{StateMatrix}
        \begin{bmatrix}
        \Delta \dot{x}_{1}  \\
        \vdots  \\
        \Delta \dot{x}_{M}  \\
        \Delta \dot{x}_{E}
        \end{bmatrix}
        =
        \underbrace{\begin{bmatrix}
            O & \cdots & R & V \\
            \vdots & \ddots & \vdots & \vdots \\
            R & \cdots & O & V \\
            S & \cdots & S & T
        \end{bmatrix}}_{A}
        \begin{bmatrix}
        \Delta x_{1}  \\
        \vdots  \\
        \Delta x_{M}  \\
        \Delta x_{E}
        \end{bmatrix}
        , \text{where}
        \left\{
        \begin{aligned}
             O&=A_{\alpha}\!+\!B_{\alpha}/(I\!-\!MD_{\beta}D_{\alpha})D_{\beta}C_{\alpha}, R=B_{\alpha}/(I\!-\!MD_{\beta}D_{\alpha})D_{\beta}C_{\alpha} \\
             S&=B_{\beta}/(I\!-\!MD_{\alpha}D_{\beta})C_{\alpha}, T=A_{\beta}\!+\!MB_{\beta}/(I\!-\!MD_{\alpha}D_{\beta})D_{\alpha}C_{\beta}\\
             V&=B_{\alpha}/(I\!-\!MD_{\beta}D_{\alpha})C_{\beta}
        \end{aligned}
        \right.
    \end{equation}
\end{figure*}

\begin{figure}[t]
\centering
\includegraphics[scale=0.87]{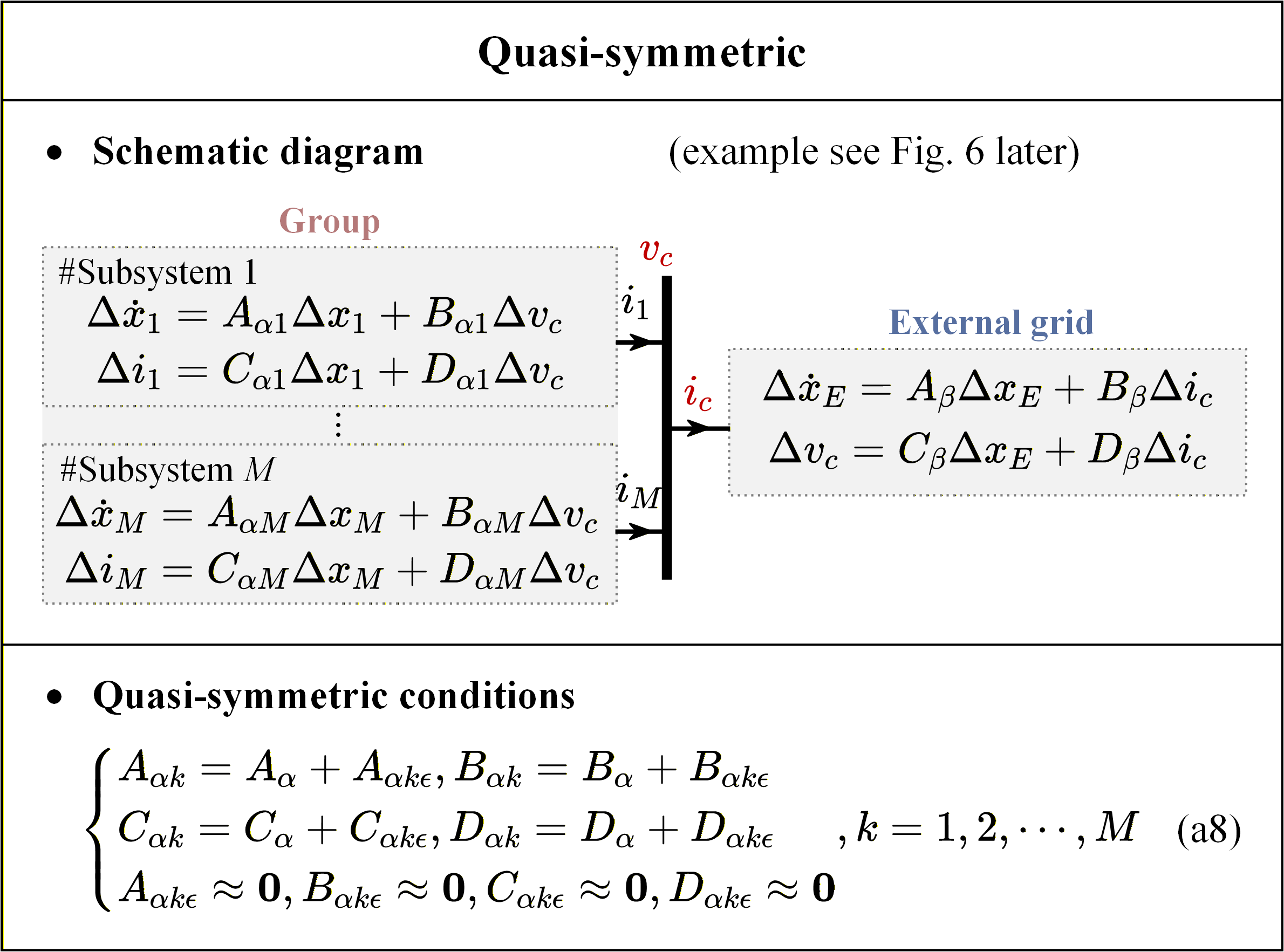}
\caption{Schematic diagram of quasi-symmetric systems.}
\label{Fig:QuasiSymmetric_EigenvaluePattern}
\end{figure}

2) \emph{Participation Factors}:
The participation factors can be computed directly from the original state matrix of the entire grid-connected system. By combining (a2) and (a3), the original state matrix $A$ is obtained as \eqref{RLStateMatrix} at the bottom of the previous page. The participation factor of the $k$-th state variable with respect to the $i$-th eigenvalue is defined as $pf_{ki}=\partial {\lambda_i}/\partial {a_{kk}}=\Psi_{ik}\Phi_{ki}$, where $a_{kk}$ is the $k$-th diagonal element of the state matrix $A$, and $\Psi_{ik}$ and $\Phi_{ki}$ denote the $(i,k)$-th and $(k,i)$-th element of the left and right eigenvector matrices $\Psi$ and $\Phi$, respectively, both corresponding to the $i$-th eigenvalue\cite{kundur1994power}. This definition holds under most operating conditions but may become invalid in systems with repeated modes, i.e., inner-group modes $\lambda_{1}$ and $\lambda_{2}$. Specifically, when the geometric multiplicity is greater than $1$ for repeated modes, the associated characteristic subspace is multi-dimensional, rendering conventional participation factors ill-defined and non-unique\cite{kundur1994power,strang2016introduction}. A detailed discussion of this issue is provided in \appendixref{Section:Conventional Participation Factors for Repeated Modes}, and its solution is presented in \sectionref{Section: Group Participation Factor}.

Nevertheless, conventional participation factors are well-defined and unique for the single, distinct group-grid modes, $\lambda_{3}$ and $\lambda_{g}$. The results are as follows:
\setcounter{equation}{4}
\begin{equation}
    \begin{aligned}
    \lambda_3:\;pf_{13}&=pf_{23}=pf_{33}\neq 0,\; pf_{g3}\neq 0\\
    \lambda_g:\;pf_{1g}&=pf_{2g}=pf_{3g}\neq 0,\; pf_{gg}\neq 0,
    \end{aligned}
\end{equation}
where the first subscript identifies the current state variable, with subscripts $1$, $2$, $3$, and $g$ corresponding to $\Delta i_{1dq+}$, $\Delta i_{2dq+}$, $\Delta i_{3dq+}$, and $\Delta i_{gdq+}$, respectively. The second subscript indicates the associated eigenvalue, where $3$ corresponds to $\lambda_3$ and $g$ to $\lambda_g$. Taking $\lambda_3$ as an example, the participation factors of $\Delta i_{1dq+}$, $\Delta i_{2dq+}$ and $\Delta i_{3dq+}$ to $\lambda_3$ are equivalent, a consequence of the identical $RL$ branch parameters. The non-zero participation factor of $\Delta i_{gdq+}$ to $\lambda_3$ indicates that this eigenvalue is affected by both the $RL$ group and the external grid, adhering to the definition of group-grid modes. A parallel analysis applies to $\lambda_g$ and thus omitted for brevity.

\subsection{General Derivation}\label{Section: General Derivation}
A generalized representation of ideally-symmetric systems is shown in the right column of \figref{Fig:IdeallySymmetric_StateSpace}. The definition is given from the perspectives of physical characteristics and the state space model as follows. The system is defined as ideally-symmetric if the group consists of $M$ subsystems with identical structure and parameters. In other words, all subsystems are described by the same state space models, which satisfy the ideally-symmetric condition (b1). This condition indicates that the state matrices of all subsystems $A_{\alpha k}, k=1,2,\cdots,M$ are equal and can be uniformly denoted as $A_{\alpha}$. The same holds for $B_{\alpha k}$, $C_{\alpha k}$, and $D_{\alpha k}$.

The state space representations of these subsystems are given in (b2). All subsystems are connected in parallel to the external grid, of which the state space representations are expressed as (b3). Similar to (a4), we perform the transformation matrix $P$ (b4) on state variables. As a result, (b2) is reformulated as (b5) and (b6), while (b3) remains unchanged. From the derived state space models, the eigenvalue and participation factor analyses are conducted next.

1) \emph{Eigenvalue Patterns}: The eigenvalues of (b5) are further calculated as
\begin{equation}
    \label{InnerMode}
    \Lambda_k=\text{eig}(A_{\alpha}), k=1,2,\cdots,M-1
\end{equation}
where $\Lambda_k$ is an $m$ order diagonal matrix. In \eqref{InnerMode}, there exist $m$ types of repeated modes, with each type of mode repeated $M-1$ times. These modes can be defined as inner-group modes of the entire grid-connected system, as (b5) does not contain the input variable $\Delta v_{c}$.

In contrast, the state space representations of (b6) involve interactions with the external grid via the input variable $\Delta v_{c}$, and need to be combined with (b3) to derive the remaining modes of the entire system, as expressed in (b7). From (b7), the eigenvalues are further calculated as
\begin{equation}
    \label{OuterMode}
    \begin{bmatrix}
        \Lambda_{M}& \\
         & \Lambda_{E}
    \end{bmatrix}
    =\text{eig}\left (
    \begin{bmatrix}
        A_{\gamma11} & A_{\gamma12} \\
        A_{\gamma21} & A_{\gamma22}
    \end{bmatrix}\right )
    =\text{eig}(A_{\gamma})
\end{equation}
where $\Lambda_{M}$ and $\Lambda_{E}$ are diagonal matrices of order $m$ and $e$, respectively. These eigenvalues can be defined as group-grid modes, as they are determined by both the group and the external grid. There are two key parameters: $c$ and $M$. $c$ is the model modification factor. As shown in (b7), it changes the expression form of the model but does not alter the eigenvalues. $M$ is the number (or scale) of subsystems within the group. It is worth noting that $M$ impacts the group-grid modes but has no influence on the inner-group modes.

2) \emph{Participation Factors}: The participation analysis in \sectionref{Section: Simple Example} is extended here. Similarly, (b2) and (b3) are combined to derive the original state matrix of the whole grid-connected system as \eqref{StateMatrix} at the bottom of this page. The corresponding participation factors are organized in a block matrix form as
\setcounter{equation}{8}
\begin{equation}
    \begin{aligned}
    &PF=\\
        &\begin{bmatrix}
            PF_{11} & \!\!\cdots & \!\!PF_{1M-1} & \!\!PF_{1M} & \!\!PF_{1E}\\
            \vdots & \!\!\ddots & \!\!\vdots & \!\!\vdots & \!\!\vdots\\
             PF_{M-11} & \!\!\cdots &  \!\!PF_{M-1M-1} &  \!\!PF_{M-1M} &  \!\!PF_{M-1E}\\
             PF_{M1} & \!\!\cdots &  \!\!PF_{MM-1} &  \!\!PF_{MM} &  \!\!PF_{ME}\\
             PF_{E1} & \!\!\cdots &  \!\!PF_{EM-1} &  \!\!PF_{EM} &  \!\!PF_{EE}
        \end{bmatrix}
    \end{aligned}
\end{equation}
where the row indices (first subscript) correspond to state variables associated with the $M$ identical subsystems ($1$ through $M$) and the external grid ($E$), while the column indices (second subscript) correspond to the system eigenvalues: the inner-group modes ($1$ through $M-1$), the group-grid modes ($M$ and $E$). The conventional participation factors for the repeated inner-group modes (column $1$ through $M-1$) remain ill-defined, as discussed in \sectionref{Section: Simple Example}. The solution to this limitation is addressed in \sectionref{Section: Group Participation Factor} through the proposed method.

For the group-grid modes $\Lambda_{M}$ and $\Lambda_{E}$, the participation factors satisfy
\begin{equation}
    \label{PF_IdeallySymmetric}
    \begin{aligned}
        PF_{1M}\!&=\cdots=\!PF_{M-1M}\!=\!PF_{MM}\!\neq \!\mathbf{0},\;PF_{EM}\!\neq \!\mathbf{0}\\
        PF_{1E}\!&=\cdots=\!PF_{M-1E}\!=\!PF_{ME}\!\neq \!\mathbf{0},\;PF_{EE}\!\neq \!\mathbf{0}.
    \end{aligned}
\end{equation}
The equality $PF_{1M}\!=\cdots=\!PF_{MM}$ indicates that all identical subsystems within the group participate equally in mode $\Lambda_{M}$, reflecting the symmetry of the system. Similarly, the equal participation factors $PF_{1E}\!=\cdots=\!PF_{ME}$ for mode $\Lambda_{E}$ also confirm this symmetric property. Furthermore, the non-zero terms $PF_{EM}$ and $PF_{EE}$ verify that both modes involve interactions with the external grid, thereby satisfying the fundamental definition of group-grid modes as eigenvalues influenced by both the group and external grid dynamics.


\section{Nonideally-Symmetric Systems} \label{Section:Nonideally-Symmetric Systems}
Ideal symmetry rarely exists in practical power systems. Instead, real-world systems often exhibit approximate symmetry, referred to as nonideally-symmetric systems. In this section, we investigate two representative classes of nonideally-symmetric renewable energy power systems: quasi-symmetric systems and group-symmetric systems.

\subsection{Quasi-Symmetric Systems}\label{Section:Quasi-Symmetric Systems}
In practice, while the internal structure of subsystems may be identical, their parameters (especially operational parameters) are often different. On this basis, a system is defined as quasi-symmetric if the $M$ subsystems within the group have identical structure and similar parameters. \figref{Fig:QuasiSymmetric_EigenvaluePattern} gives the corresponding quasi-symmetric condition (a8), which is derived by modeling the quasi-symmetric system as a small perturbation of an ideally-symmetric one. For instance, the state matrix of each subsystem is expressed as $A_{\alpha k}=A_{\alpha}+A_{\alpha k \epsilon},k=1,2,\cdots,M$, where $A_{\alpha k \epsilon}$ represents a small perturbation matrix. When these perturbations are sufficiently small, i.e., $A_{\alpha k \epsilon} \approx \mathbf{0}$, the system qualifies as quasi-symmetric. The same holds for $B_{\alpha k}$, $C_{\alpha k}$, and $D_{\alpha k}$.

The state space representations of the group are expressed as
\begin{equation}
    \begin{aligned}
        \Delta \dot{x}_{k}&=A_{\alpha k} \Delta x_{k}+B_{\alpha k} \Delta v_{c}, k=1,2, \cdots, M\\
        \Delta i_{c}&=\sum_{k=1}^{M} \Delta i_{k}=\sum_{k=1}^{M} C_{\alpha k}\Delta x_{k}+\sum_{k=1}^{M} D_{\alpha k}\Delta v_{c}.
    \end{aligned}
\end{equation}
Similarity transformation is implemented with (b4), and the equivalent state space representations of the group are derived as
\begin{equation}
    \begin{aligned}
        \Delta \dot{x}_{k}^{\prime}&=A_{\alpha} \Delta x_{k}^{\prime}+\sum_{j=1}^{M}p_{kj}A_{\alpha j \epsilon}\Delta x_{j}+\sum_{j=1}^{M}p_{kj}B_{\alpha j \epsilon}\Delta v_{c}\\
        &\approx A_{\alpha} \Delta x_{k}^{\prime},k=1,2, \cdots, M-1
    \end{aligned}
    \label{QuasiSymmetricStateSpaceM-1}
\end{equation}
and
\begin{equation}
    \begin{aligned}
        \Delta \dot{x}_{M}^{\prime}\!&=\!A_{\alpha} \Delta x_{M}^{\prime}\!+\!cM\!B_{\alpha} \Delta v_{c}\!+\!c\!\sum_{k=1}^{M}\!A_{\alpha k \epsilon}\Delta x_{k}\!+\!c\!\sum_{k=1}^{M}\!B_{\alpha k \epsilon}\Delta v_{c}\\
        &\approx\!A_{\alpha} \Delta x_{M}^{\prime}\!+\!cM\!B_{\alpha} \Delta v_{c}\\
        \Delta i_{c}\!&=\!\frac{1}{c}C_{\alpha}\Delta x_{M}^{\prime}\!+\!M\!D_{\alpha}\Delta v_{c}\!+\!\sum_{k=1}^{M}\!C_{\alpha k \epsilon}\Delta x_{k}\!+\!\sum_{k=1}^{M}\!D_{\alpha k \epsilon}\Delta v_{c}\\
        &\approx\!\frac{1}{c}C_{\alpha}\Delta x_{M}^{\prime}\!+\!M\!D_{\alpha}\Delta v_{c}.
    \end{aligned}
    \label{QuasiSymmetricStateSpaceM}
\end{equation}
It is obvious that the structure of \eqref{QuasiSymmetricStateSpaceM-1} and \eqref{QuasiSymmetricStateSpaceM} is similar to that of (b5) and (b6), with the main difference being the use of approximate equality ($\approx$) instead of exact equality ($=$). Therefore, the modal analysis results of ideally-symmetric systems can be extended to quasi-symmetric systems as follows.

1) \emph{Eigenvalue Patterns}: From \eqref{QuasiSymmetricStateSpaceM-1}, the following approximation holds:
\begin{equation}
    \label{QuasiSymmetricInnerMode}
    \Lambda_{k,\text{quasi}}\approx \Lambda_{k,\text{ideal}}, k=1,2,\cdots,M-1
\end{equation}
where $\Lambda_k, k=1,2,\cdots,M-1$ denotes a subset of the actual eigenvalues of the grid-connected systems. Compared with \eqref{InnerMode}, there are $M-1$ close modes for each type in \eqref{QuasiSymmetricInnerMode}, split from $M-1$ repeated modes in the ideally-symmetric system. These modes can still be regarded as inner-group modes, due to the little influence from the external grid. The remaining eigenvalues are approximately calculated as
\begin{equation}
    \label{QuasiSymmetricOuterMode}
    \begin{bmatrix}
        \Lambda_{M,\text{quasi}}& \\
         & \Lambda_{E,\text{quasi}}
    \end{bmatrix}
    \approx
    \begin{bmatrix}
        \Lambda_{M,\text{ideal}}& \\
         & \Lambda_{E,\text{ideal}}
    \end{bmatrix},
\end{equation}
which are still defined as group-grid modes, as they are affected by both the group and the external grid.

2) \emph{Participation Factors}: It is pointed out in \sectionref{Section:Ideally-Symmetric Systems} that conventional participation factors for inner-group (repeated) modes are invalid. In quasi-symmetric systems, the inner-group modes are close modes, for which the participation factors are sensitive, as proven in the literature\cite{shao2023participation}. This means that even a slight perturbation in the parameters is likely to introduce significant differences to participation factors for close modes. The sensitivity is not conducive to accurately assessing the contribution of each subsystem to the modes (weak robustness). This limitation is also solved in \sectionref{Section: Group Participation Factor}.

For the group-grid modes $\Lambda_{M}$ and $\Lambda_{E}$, the participation factors satisfy
\begin{equation}
    \label{PF_QuasiSymmetric}
    \begin{aligned}
        PF_{1M}\!&\approx\cdots\approx\!PF_{M-1M}\!\approx\!PF_{MM}\!\neq \!\mathbf{0},\;PF_{EM}\!\neq \!\mathbf{0}\\
        PF_{1E}\!&\approx\cdots\approx\!PF_{M-1E}\!\approx\!PF_{ME}\!\neq \!\mathbf{0},\;PF_{EE}\!\neq \!\mathbf{0}.
    \end{aligned}
\end{equation}
Compared with \eqref{PF_IdeallySymmetric} in the ideally-symmetric system, \eqref{PF_QuasiSymmetric} indicates that in the quasi-symmetric system, all subsystems within the group participate nearly equally in both modes, thereby preserving the symmetric characteristic. Furthermore, the non-zero terms $PF_{EM}$ and $PF_{EE}$ verify that both modes interact with the external grid, which also satisfies the fundamental definition of group-grid modes.

\subsection{Group-Symmetric Systems}
\begin{figure}[t]
\centering
\includegraphics[scale=0.87]{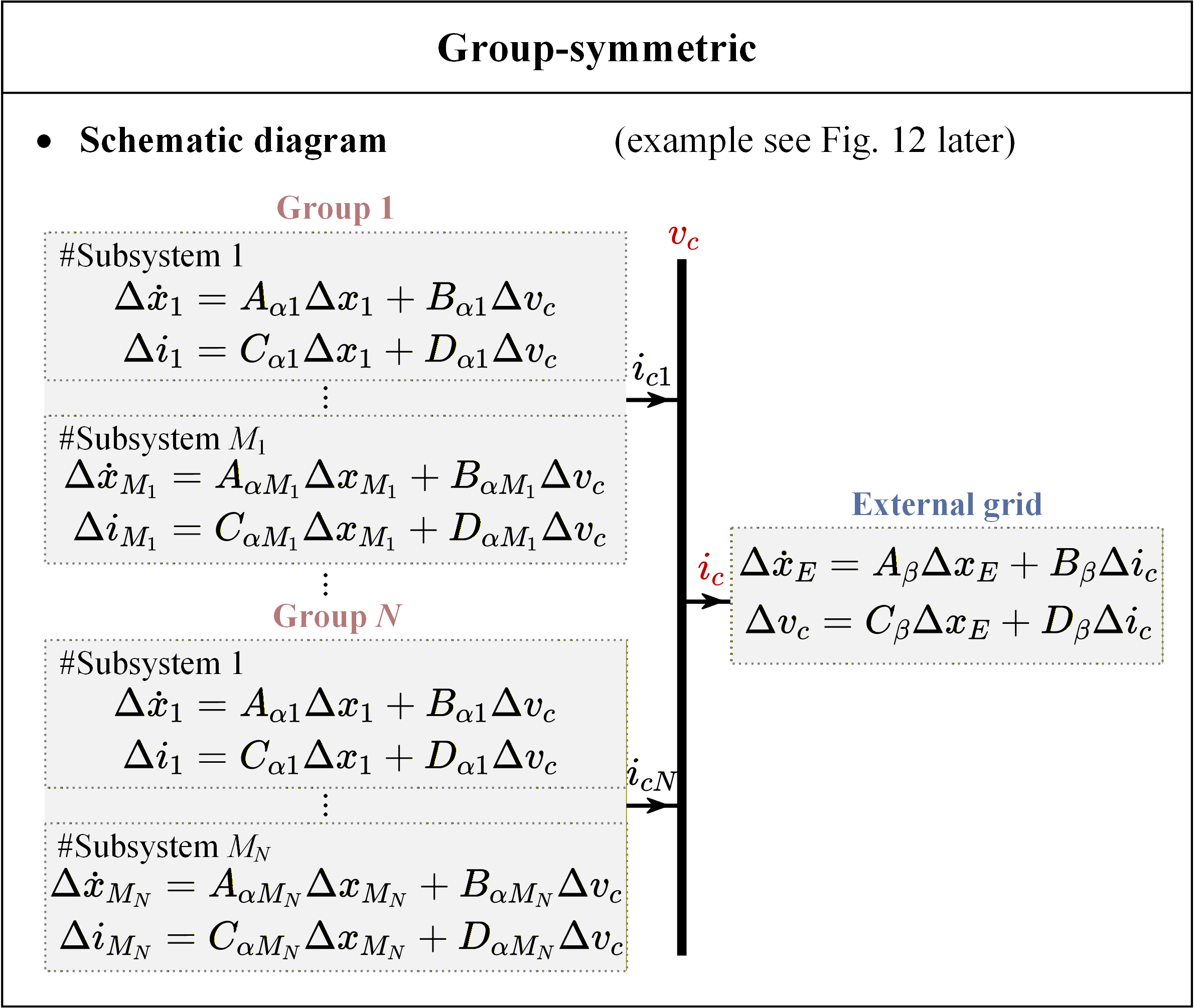}
\caption{Schematic diagram of group-symmetric systems.}
\label{Fig:GroupSymmetric_EigenvaluePattern}
\end{figure}

In the ideally- and quasi-symmetric systems discussed previously, all subsystems belong to a single group with identical structure and either identical (ideally-symmetric) or similar (quasi-symmetric) parameters. However, in renewable energy power systems, generation units are not necessarily homogeneous, particularly when they originate from different manufacturers. Consequently, the subsystems may not satisfy the ideally- or quasi-symmetric conditions. To address this, the system can be partitioned into multiple groups. As depicted in \figref{Fig:GroupSymmetric_EigenvaluePattern}, subsystems with identical structure and identical (or similar) parameters are classified into the same group, and the entire system is composed of $N$ such different groups. Each group can be ideally-symmetric or quasi-symmetric, based on whether its internal subsystems have identical or similar parameters, respectively. Therefore, the eigenvalue patterns and participation factors of group-symmetric systems can be derived by building upon the analysis developed for ideally-symmetric and quasi-symmetric systems.

1) \emph{Eigenvalue Patterns}: For ease of analysis, we first consider the simplest case in which all groups are assumed to be ideally-symmetric. The similarity transformation is implemented, and the equivalent state space representations of the group with the form of (b5) and (b6) is derived. For an arbitrary group indexed by $j=1,2,\cdots,N$, the eigenvalues of (b5) are inner-group modes, which are only affected by the subsystems within the $j$-th group. On the other hand, the state space representations of (b6) across all groups are merged with those of the external grid (b3), forming the rest state space model of the grid-connected system. The corresponding eigenvalues represent group-grid modes, which are affected both by each group and the external grid. It is noteworthy that similar conclusions hold if some or all groups become quasi-symmetric, as demonstrated in \sectionref{Section:Quasi-Symmetric Systems}.

2) \emph{Participation Factors}: In such multi-group systems, the conventional participation factors for the inner-group modes within each symmetric group encounter the same limitations as discussed before. Specifically, these participation factors become ill-defined for repeated modes in ideally-symmetric groups and exhibit high sensitivity for close modes in quasi-symmetric groups. In contrast, the group-grid modes of the overall system possess well-defined participation factors. For any group-grid mode, the participation factors of subsystems belonging to the same group show equal or approximately equal values, yet distinct from those of other groups. Furthermore, these participation factors exhibit non-zero participation from external grid states.

\section{Extension and Application of Symmetry}\label{Section: Extension and Application of Symmetry}
\subsection{Group Participation Factor}\label{Section: Group Participation Factor}
As mentioned in \sectionref{Section:Ideally-Symmetric Systems} and \sectionref{Section:Nonideally-Symmetric Systems}, conventional participation factors are ineffective for analyzing inner-group modes. To avoid this problem, a new concept termed group participation factor is proposed in this paper. This section will present a theoretical demonstration of its advantages and provide the corresponding results for inner-group modes.

First, for ideally-symmetric systems, the group participation factor is defined as the collective influence of the $k$-th state variable on the set of repeated modes:
\begin{equation}
    \label{GPF_IdeallySymmetric}
    gpf_{k}=\sum_{i=1}^{n_g}\Psi_{ik}\Phi_{ki},
\end{equation}
where $\lambda_i,i=1,\cdots,n_g$ are the repeated modes and $n_g$ is their algebraic multiplicity. Given that the geometric multiplicity equals the algebraic multiplicity, a complete set of $n_g$ linearly independent eigenvectors exists. Consequently, \eqref{GPF_IdeallySymmetric} sums the $n_g$ different participation factors corresponding to the same repeated mode. Based on this definition, the group participation factor is always well-defined and unique, as demonstrated in \appendixref{Section:Group Participation Factors for Repeated Modes}.

This concept can now be applied to address the problem left in the previous simple example in \sectionref{Section: Simple Example}. For the state matrix in \eqref{RLStateMatrix}, the group participation factors for the inner-group modes satisfy
\begin{equation}
    \label{GPF_InnerModes12}
    \{\lambda_1,\lambda_2\}:\;gpf_{1}=gpf_{2}=gpf_{3}\neq 0,\;
        gpf_{g}=0.
\end{equation}
As shown in \eqref{GPF_InnerModes12}, the group participation factors of the state variables $\Delta i_{1dq+}$, $\Delta i_{2dq+}$, and $\Delta i_{3dq+}$ to \{$\lambda_1,\lambda_2$\} are identical. It is reasonable due to the symmetry of the three $RL$ branches. The group participation factor of the external grid state variable $\Delta i_{gdq+}$ is zero, consistent with the definition of inner-group modes.

Similarly, for the general case derived in \sectionref{Section: General Derivation}, the group participation factors of the inner-grid modes \{$\Lambda_1,\Lambda_2,\cdots,\Lambda_{M-1}$\} satisfy
\begin{equation}
    \label{GPF_E}
    GPF_{1}\!=\!\cdots\!=\!GPF_{M-1}\!=\!GPF_{M}\!\neq\! \mathbf{0},\;GPF_{E}\!=\!\mathbf{0},
\end{equation}
where each $GPF_{k},k=1,2,\cdots,M-1,M,E$ is computed as
\begin{equation}
    GPF_{k}=PF_{k1}+PF_{k2}+\cdots+PF_{kM-1}.
\end{equation}

Second, for quasi-symmetric systems, the group participation factor is extended to analyze closely spaced modes, replacing the conventional participation factor. The formulation remains identical to \eqref{GPF_IdeallySymmetric}, with the distinction that $n_g$ now represents the number of close modes of the same type. Based on this definition, the group participation factors for close modes are robust (insensitive) to parameter variations, as demonstrated in \appendixref{Section:Group Participation Factors for Close Modes}.

This concept is then applied to address the problem left in the quasi-symmetric system presented in \sectionref{Section:Quasi-Symmetric Systems}. For the set of inner-grid modes \{$\Lambda_1,\Lambda_2,\cdots,\Lambda_{M-1}$\}, the analysis yields
\begin{equation}
    GPF_{1}\!\approx\!\cdots\!\approx\!GPF_{M-1}\!\approx\!GPF_{M}\!\neq\! \mathbf{0},\;GPF_{E}\!\approx\!\mathbf{0}.
\end{equation}

The above participation analysis not only reflects the physical significance of the inner-group modes but also conforms to the intuitive understanding of symmetric subsystems.

\subsection{Invariance Properties of Modes}\label{Section: Invariance Properties of Modes}
In the previous work, we conducted an in-depth analysis of the eigenvalue patterns and participation factors in renewable energy power systems based on symmetry. To strengthen the theoretical framework, the relationship between symmetry and conservation in physics is introduced into the subsequent analysis.

In the case of an ideally-symmetric system, the inner-group modes calculated from (b5) exhibit a zero group participation factor for the external grid, i.e., $GPF_{E}=\mathbf{0}$ as given in \eqref{GPF_E}, due to the exclusion of the input term $\Delta v_c$. As a result, these modes remain nearly invariant despite changes in the external grid. This property highlights the inherent independence of group's internal dynamics from external grid conditions under ideal symmetry. On the other hand, from (b6) we can see that the input and output variables are $\Delta v_c$ and $\Delta i_c$ at the point of common coupling (PCC), respectively. The input-output relationship at the PCC describes the terminal behavior of the entire group. Consequently, the group-grid modes capture the dynamic interactions between the aggregated group and the external grid. These modes are almost unaffected by internal changes within the group, provided that the group's terminal dynamics remain constant. Specifically, if the parameters of a few subsystems within the group are slightly changed, the group-grid modes remain nearly invariant because such changes have a small impact on the group's overall dynamics.

The invariance properties provide two key insights. First, the invariant inner-group modes indicate that stabilizing these modes is most effectively achieved by adjusting subsystems within the associated group, rather than modifying other parts of the system. Furthermore, adjusting only a few subsystems is insufficient. It is recommended to tune all subsystems within the group simultaneously to shift the corresponding inner-group modes leftward in the complex plane. Second, the invariant group-grid modes demonstrate that making slight changes to a few subsystems within a group has little impact on these modes, as the terminal characteristics seen from the PCC remain largely unchanged.

There are a few points that need to be clarified. Firstly, the symmetry is not the only factor determining this invariance property. Another key factor is the location of the parameter variation. The observed invariance is conditional and relative to specific types of perturbations. For instance, while the inner-group modes are invariant to changes in the external grid, they are, by definition, sensitive to alterations in the internal parameters of the group's subsystems. Conversely, the group-grid modes are robust to minor internal parameter changes but will shift significantly with variations in the external grid.

Secondly, the term ``invariance'' does not imply that the modes remain completely unchanged, but rather that their variation is relatively small. From a practical perspective, a change in the external grid can affect the steady-state operating points of group's internal subsystems, thereby indirectly influencing their dynamics, and vice versa. For evaluating this ``near-invariance'', a practical metric is the relative change, defined as
\begin{equation}
    \label{RelativeChange}
    RC_\lambda=\frac{|\lambda_a-\lambda_b|}{|\lambda_b|} \times 100\%
\end{equation}
where $\lambda_b$ and $\lambda_a$ denote the eigenvalue before and after the parameter variation, respectively. A small value of this metric indicates effective invariance. Quantitative results based on this metric are presented and discussed in the subsequent case studies (\figref{Fig:Case2_IdeallySymmetric_InvariantMode}). It will show that the relative change is less than 1\%, confirming the invariant characteristic. It should be noted that the relative change threshold for invariance is not universal, making it a topic for future work.

Finally, the previous analysis is general and, consequently, the resulting invariance property is applicable to all oscillation modes, including low, medium, and high frequency modes. Furthermore, for quasi-symmetric systems, similar invariance properties for both inner-group and group-grid modes can be derived using an analogous methodology. A detailed derivation for the quasi-symmetric case is omitted here for brevity.

\subsection{Summary and Application of Symmetry}\label{Section: Summary and Application of Symmetry}
\begin{figure}[t]
\centering
\includegraphics[scale=0.57]{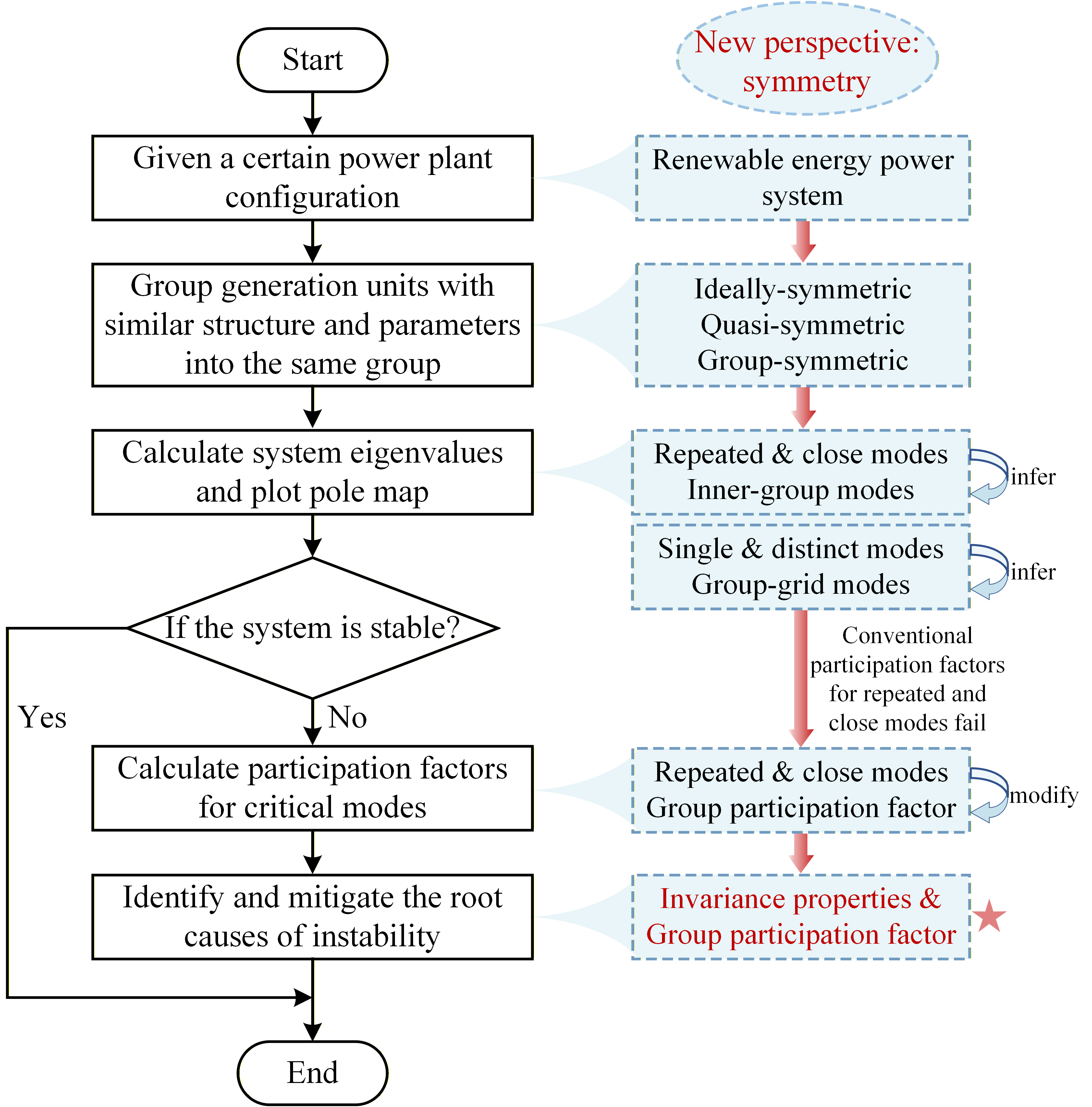}
\caption{Flow diagram illustrating how to use symmetry in a systematic way.}
\label{Fig:FlowDiagram}
\end{figure}

\figref{Fig:FlowDiagram} presents a flow diagram illustrating how to use symmetry in a systematic way for a given power plant configuration. The procedure covers all relevant aspects of eigenvalue patterns and participation factors in symmetric renewable energy power systems. After grouping generation units, the system can be classified as ideally-, quasi-, or group-symmetric configurations. The system eigenvalues are then evaluated to reveal modal patterns. Repeated and close modes are identified as inner-group modes, whereas single and well-separated modes correspond to group-grid modes. When instability is detected, conventional participation factors are inadequate for interpreting repeated or close modes. In such cases, the analysis naturally transitions to the group participation factor, which, combined with modal invariance properties, provides reliable insights into the root causes of instability. The overall framework highlights how symmetry offers a new perspective for analyzing system behavior and guiding stabilizing actions in renewable energy power systems.


\section{Case Studies} \label{Section:Case Studies}
This section validates the effectiveness of the theoretical findings through three case studies conducted in Matlab/Simulink. The parameters, scripts, and models used in case studies are open-source online\cite{casedata} and are integrated into Simplus Grid Tool\cite{simplus}. All simulation tests are based on an average model, where the nonlinear dynamics of switching are neglected. This model is valid for dynamic analysis within a bandwidth of half the switching frequency\cite{wang2019harmonic}, which is sufficient for the subsequent analysis.

Prior to the state space analysis, the dynamic models of the grid-forming (GFM) and grid-following (GFL) inverters used in the case studies are developed in the synchronous $dq$ frame. The GFM inverters are filtered by $LCL$ filters, with the inductor currents $i_{ldq}$, capacitor voltages $v_{odq}$, and output currents $i_{odq}$ as the state variables. The active power frequency droop ($P-\omega$ droop) control with a low-pass filter (LPF) is used, which outputs the frequency $\omega$ and phase angle $\delta$. The capacitor voltage $v_{odq}$ is regulated by a dual-loop voltage and current controller. The state variable for the voltage control loop is $v_{odq,i}$, and its output serves as the reference $i_{ldq}^{*}$ for the current control loop, which has $i_{ldq,i}$ as its state variable. The GFL inverters utilize $L$ filters, with the output current $i_{dq}$ as the state variable. A phase-locked loop (PLL) synchronizes the inverter with the outer grid. The current $i_{dq}$ is regulated by a current control loop, the state variable of which is $i_{dq,i}$. In the subsequent analysis, the numerical subscript on each variable corresponds to the bus number to which the inverter is connected. For example, in \sectionref{Case 1: Ideally-Symmetric System}, $\omega_2$ and $\delta_2$ denote the output frequency and phase angle of GFM2, which is connected to Bus 2.
\subsection{Case 1: Ideally-Symmetric System}\label{Case 1: Ideally-Symmetric System}
\begin{figure}[t]
\centering
\includegraphics[scale=1]{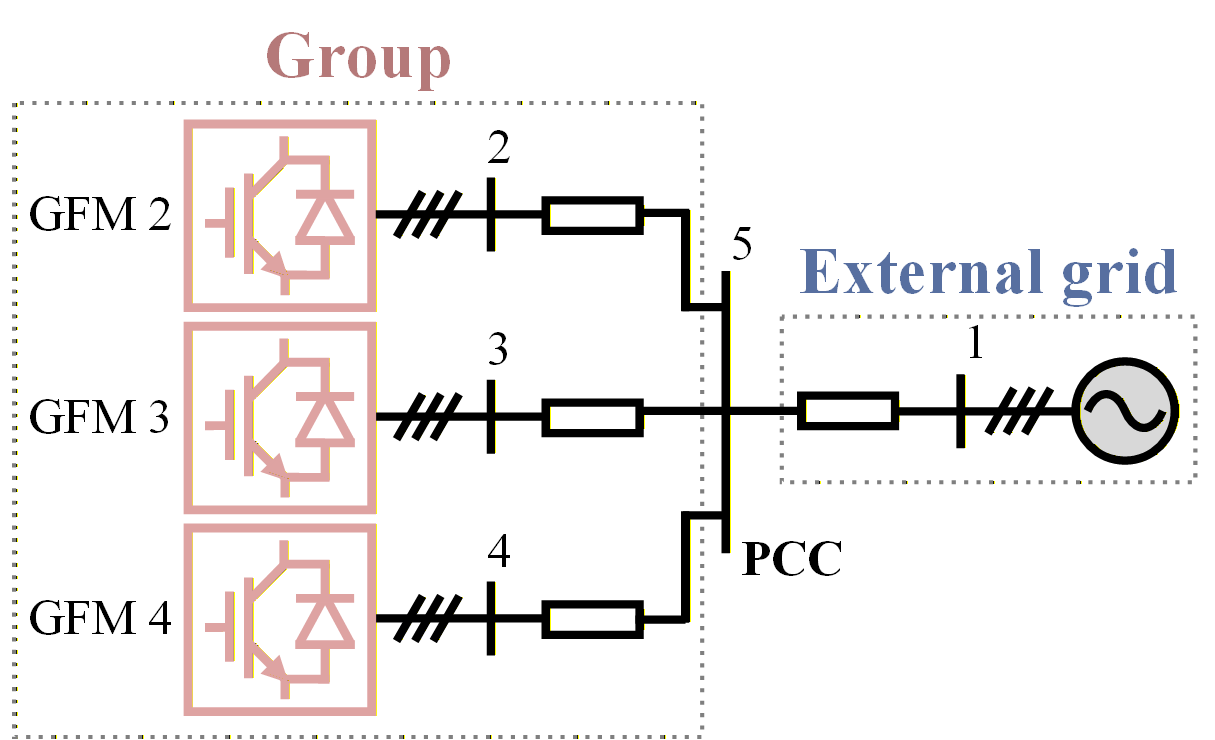}
\caption{Layout of the ideally-symmetric and quasi-symmetric system.}
\label{Fig:Case2_IdeallySymmetric_Layout}
\end{figure}

\begin{figure}[t]
\centering
\includegraphics[scale=0.44]{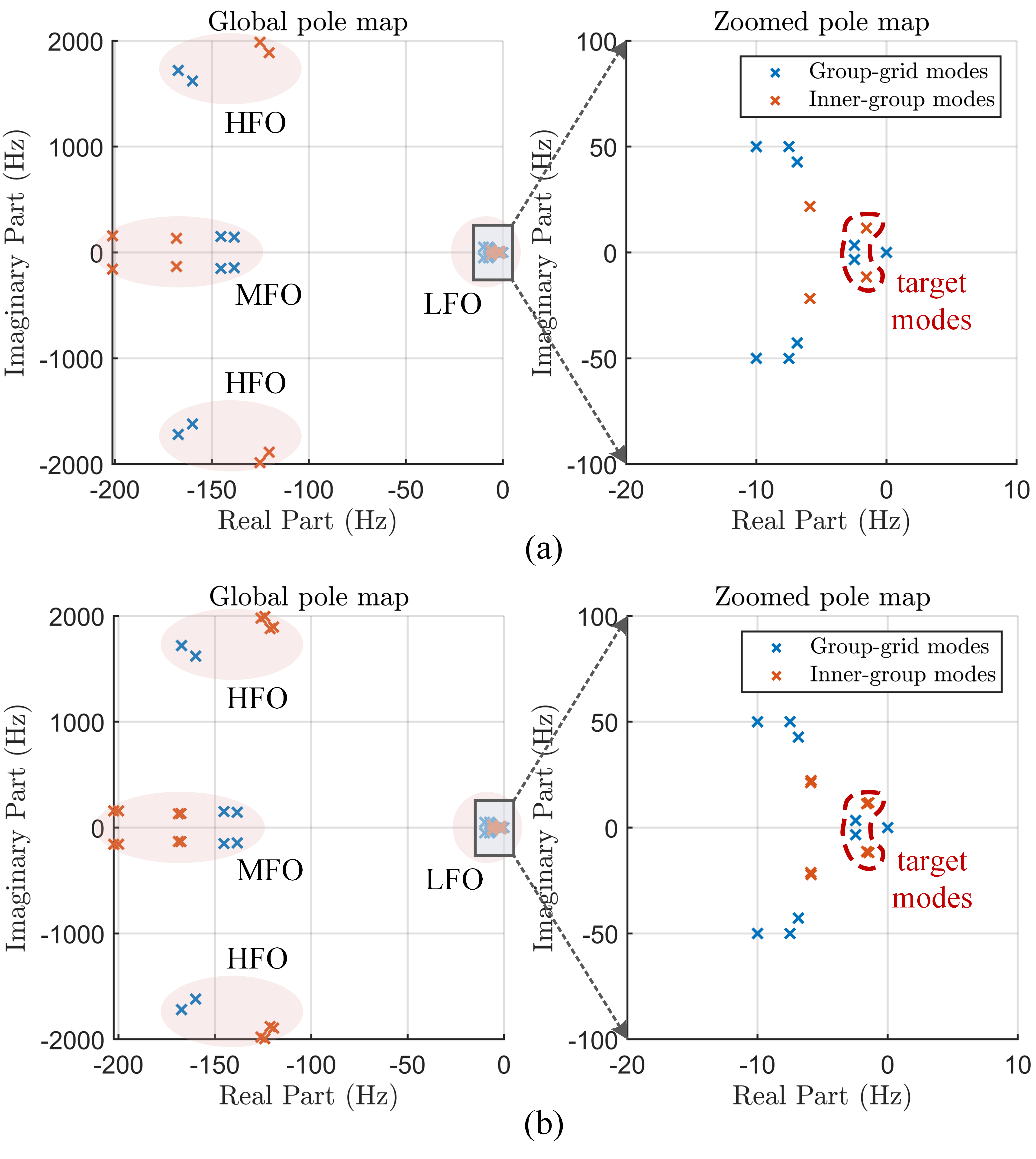}
\caption{Pole map of the whole grid-connected system. (a) Ideally-symmetric system. (b) Quasi-symmetric system.}
\label{Fig:Case23_PoleMap}
\end{figure}

\begin{figure*}[t]
\centering
\includegraphics[scale=0.36]{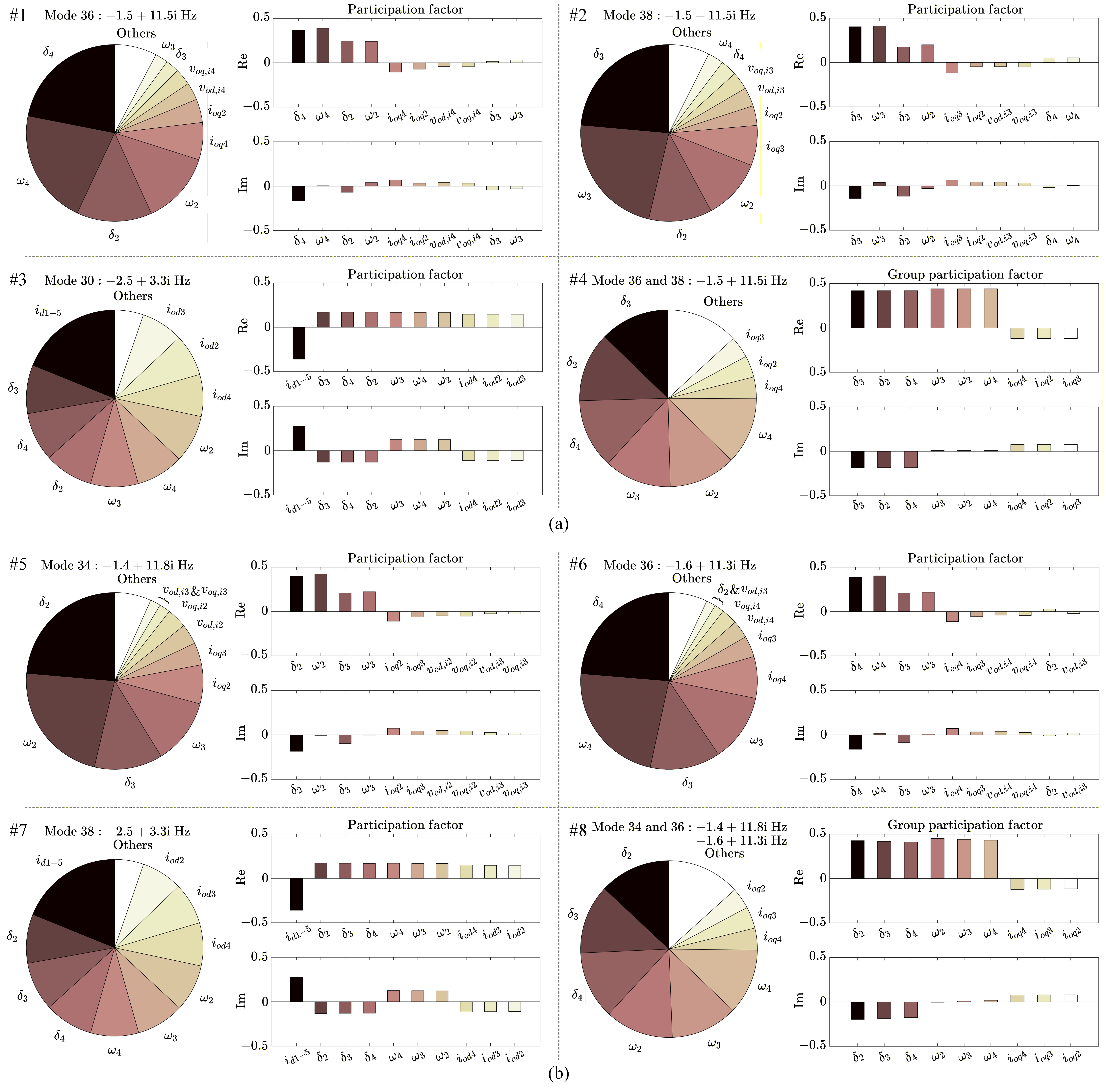}
\caption{Participation analysis of the target modes. (a) Ideally-symmetric system. (b) Quasi-symmetric system.}
\label{Fig:Case23_PF}
\end{figure*}

\begin{figure}[t]
\centering
\includegraphics[scale=0.61]{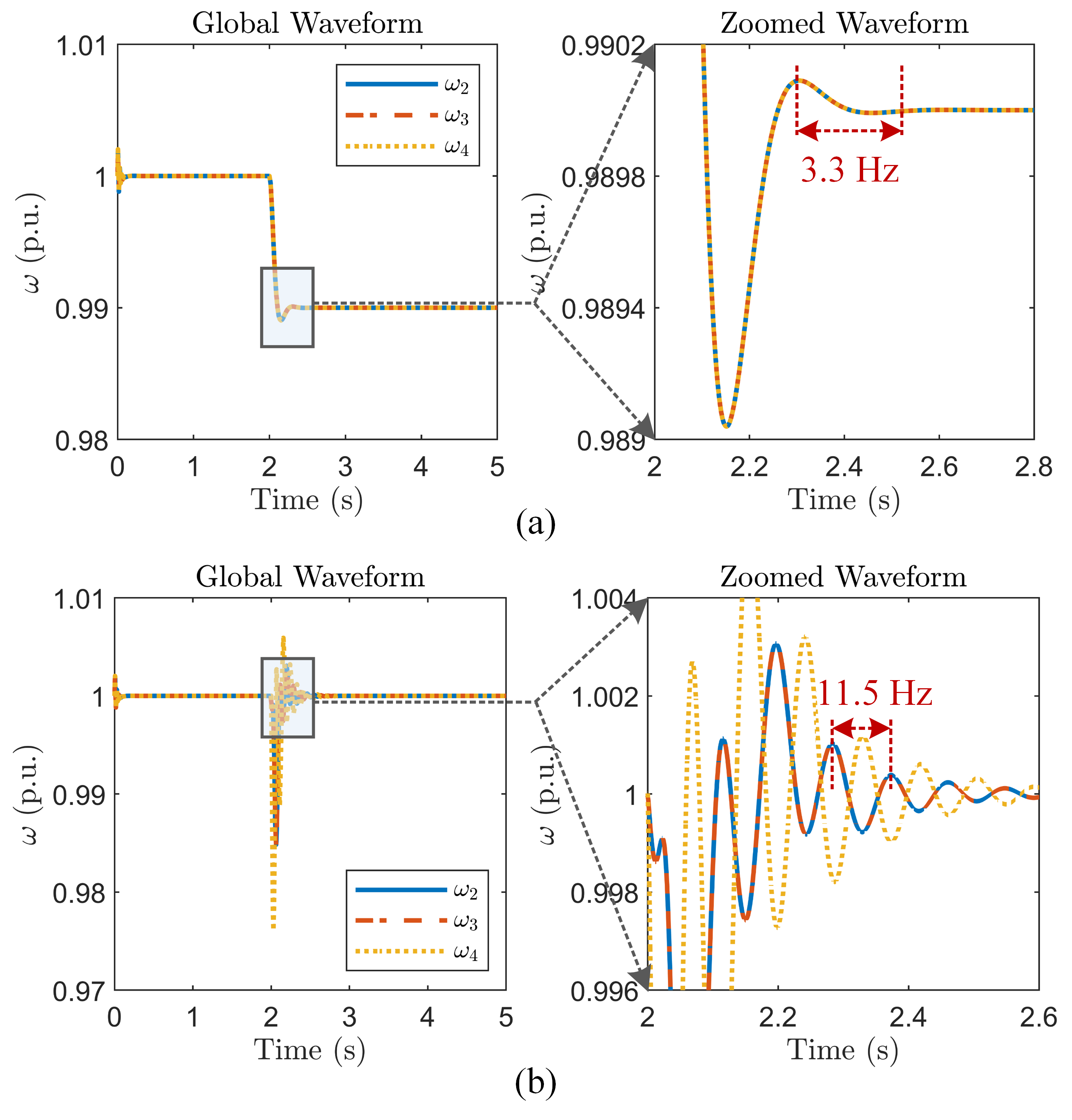}
\caption{Frequency responses at three GFM buses under different perturbations in ideally-symmetric system. (a) A disturbance changes the external grid frequency from 1 p.u. to 0.99 p.u. at 2 s. (b) A 1 p.u. resistive load step is applied at GFM4 at 2 s.}
\label{Fig:Case2_IdeallySymmetric_Simulation}
\end{figure}

\begin{figure}[t]
\centering
\includegraphics[scale=0.6]{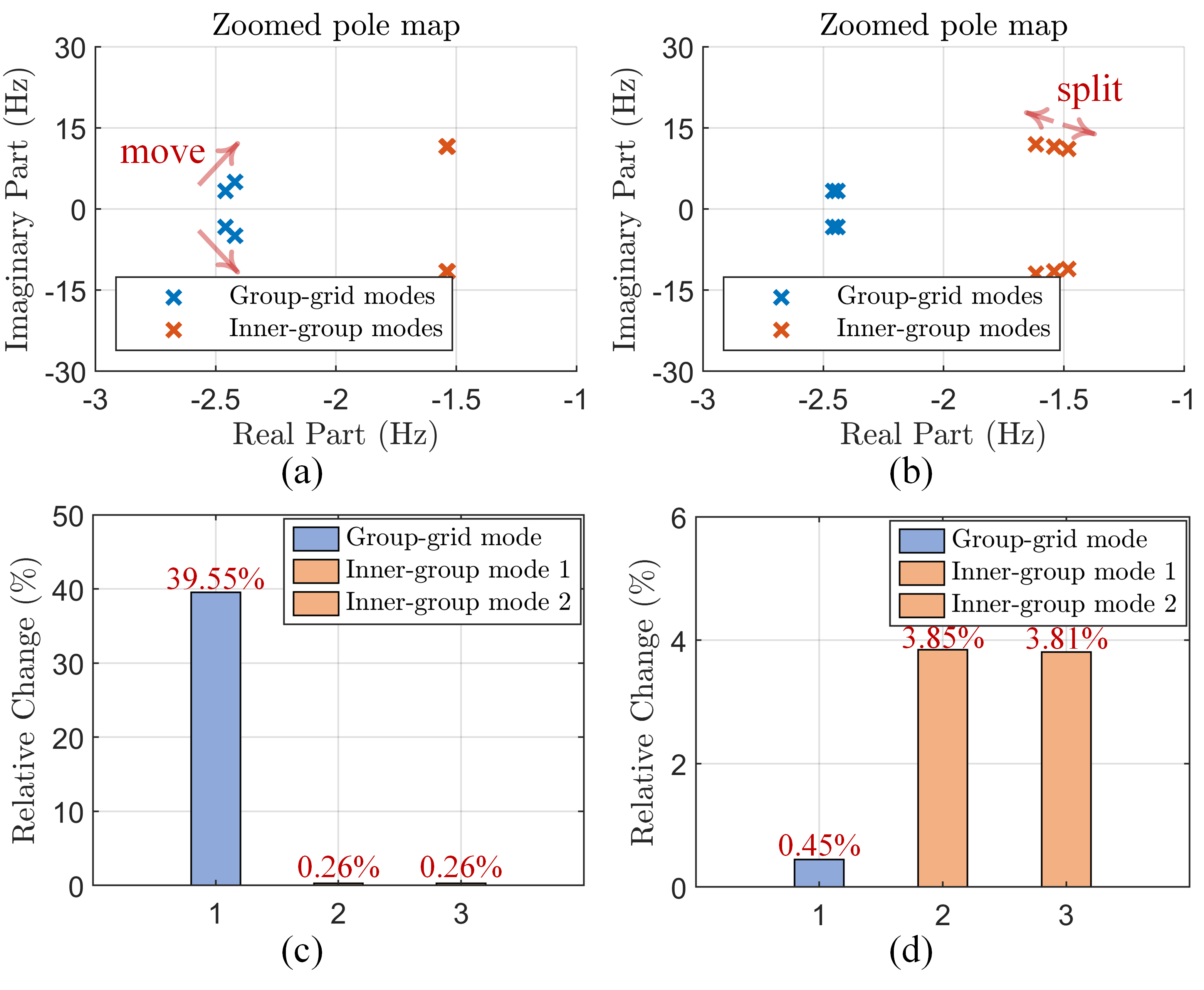}
\caption{Root loci and relative changes of the target modes under different conditions. (a) Root loci when the impedance of the external grid $Z_{1-5}$ is reduced by 50\%. (b) Root loci when frequency droop bandwidths of GFM2 and GFM4 are independently adjusted by 10\% above and below their nominal values, respectively. (c) Relative changes of the modes corresponding to (a). (d) Relative changes of the modes corresponding to (b).}
\label{Fig:Case2_IdeallySymmetric_InvariantMode}
\end{figure}

\begin{figure}[t]
\centering
\includegraphics[scale=0.3]{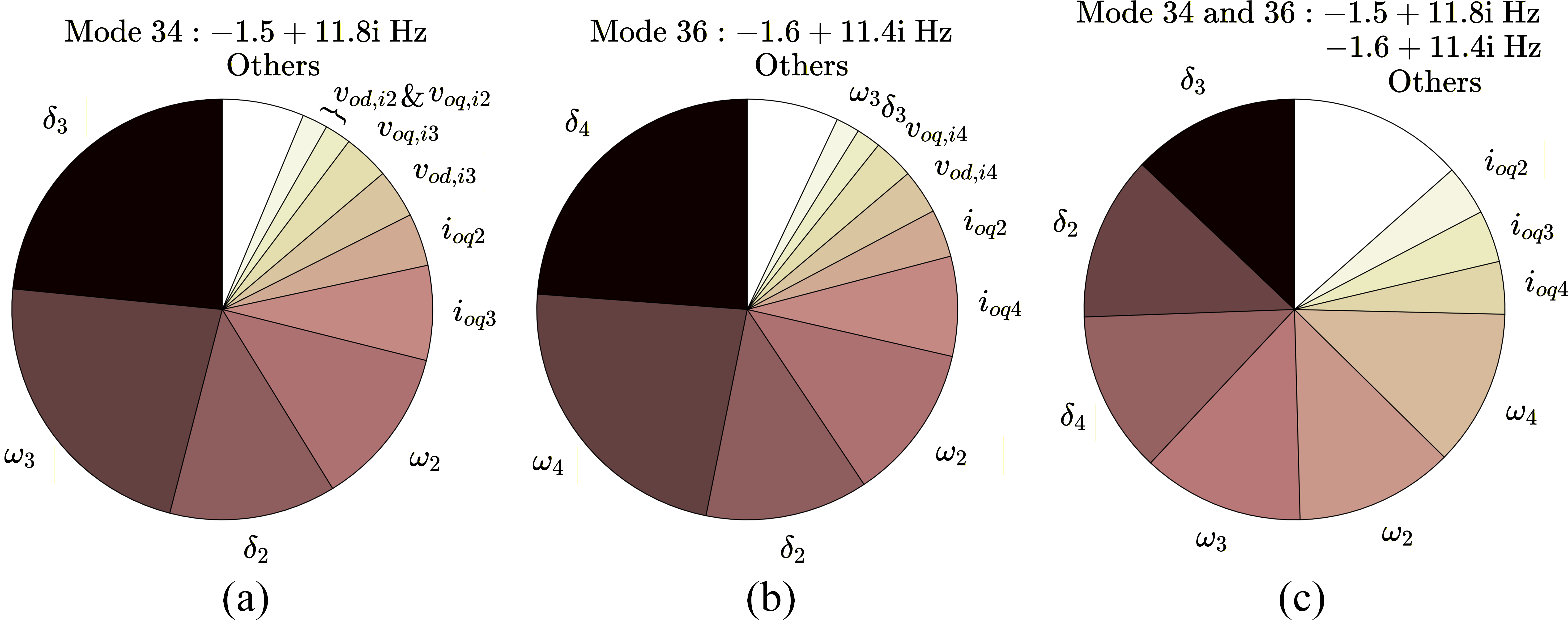}
\caption{Participation analysis of the target close modes in the quasi-symmetric system with parameter perturbations. (a) Participation factors for Mode 34. (b) Participation factors for Mode 36. (c) Group participation factors for Mode 34 and 36.}
\label{Fig:Case3_QuasiSymmetric_Perturbation}
\end{figure}

In \figref{Fig:Case2_IdeallySymmetric_Layout}, three GFM inverters are in parallel connection through line impedances to an external grid. The inverters are identical in structure and parameters, and the line impedances from each inverter to the PCC are equal. This configuration can represent a simplified, abstract model of a grid-forming energy storage system (GFM-ESS), where the remaining power system components are aggregated into the external grid. As an ideally-symmetric system, this simple case serves to clearly illustrate eigenvalue patterns and participation factor properties, as shown next.

\figref{Fig:Case23_PoleMap}(a) displays the pole map, including high frequency oscillation (HFO), medium frequency oscillation (MFO), and low frequency oscillation (LFO) modes. Among them, the LFO modes at 11.5 Hz and 3.3 Hz are of particular interest for stability analysis, as they appear on the rightmost side of the complex plane. It is worth noting that the 11.5 Hz modes are repeated, so we can naturally infer that they are inner-group modes. In contrast, the 3.3 Hz mode is a group-grid mode. The participation factors would further prove this point as follows.

\figref{Fig:Case23_PF}(a) depicts the participation pie charts and bar charts of the target modes. For the repeated 11.5 Hz modes (Mode 36 and Mode 38), the dominant participating states are internal variables of the local inverters, confirming that these modes are inner-group modes. However, it is also observed that the participation of the same state differs between Mode 36 and Mode 38, and the participation of the same type of state for a specific mode varies across the inverters. Such variations may cause confusion for operators in practical applications when interpreting participation results. To address this, the group participation factor is applied to the set of Mode 36 and Mode 38 as subfigure \#4. This mode set characterizes the dynamic interactions between the synchronization control loop ($\delta_2,\delta_3,\delta_4,\omega_2,\omega_3,\omega_4$) and the coupling inductor ($i_{oq2},i_{oq3},i_{oq4}$) of inverters.

In contrast, Mode 30 at 3.3 Hz exhibits a different participation pattern. As shown in subfigure \#3, the distribution of participation factors reveals a strong coupling between the inverter group ($\delta_2,\delta_3,\delta_4,\omega_2,\omega_3,\omega_4,i_{od2},i_{od3},i_{od4}$) and the external grid current $i_{d1-5}$. Here, the subscript ``1-5'' denotes the transmission line connecting Bus 1 and Bus 5. Therefore, the 3.3 Hz mode is not confined within the inverter group but involves significant grid dynamics, thus classifying it as a group-grid mode. The identical contributions from each inverter confirm their symmetrical involvement in the mode.

\figref{Fig:Case2_IdeallySymmetric_Simulation} shows the frequency responses of the three GFM inverters under different disturbance scenarios. In \figref{Fig:Case2_IdeallySymmetric_Simulation}(a), a frequency perturbation is introduced at 2 s on the external grid. All three inverters exhibit 3.3 Hz damped oscillations, reflecting group-grid interactions. \figref{Fig:Case2_IdeallySymmetric_Simulation}(b) applies a load step disturbance at 2 s to GFM4, and three inverters exhibit 11.5 Hz oscillations, which only occur within the inverter group.

The invariance properties of modes, as discussed in \sectionref{Section: Invariance Properties of Modes}, are further validated in \figref{Fig:Case2_IdeallySymmetric_InvariantMode}. In \figref{Fig:Case2_IdeallySymmetric_InvariantMode}(a), when the impedance of the external grid is reduced (stronger grid), the group-grid modes highlighted in blue move to the upper right of the complex plane (worse stability), confirming that multi-GFM inverter parallel grid-connected systems also suffer from strong grid instability problem like single GFM inverter cases. On the contrary, the inner-group modes highlighted in orange remain almost unaffected. In \figref{Fig:Case2_IdeallySymmetric_InvariantMode}(b), when the frequency droop control bandwidths of the GFM inverters are varied, the inner-group modes split and deviate from their repeated form, while the group-grid modes remain nearly unchanged. The relative changes of the inner-group and group-grid modes under the respective scenarios of \figref{Fig:Case2_IdeallySymmetric_InvariantMode}(a) and (b) are quantified in \figref{Fig:Case2_IdeallySymmetric_InvariantMode}(c) and (d), using the metric defined in \eqref{RelativeChange}. For conjugate mode pairs, the mode with the positive imaginary part is selected. Since the inner-group modes consist of two repeated eigenvalues, their relative changes are calculated separately, which are marked in orange color as inner-group mode 1 and 2. \figref{Fig:Case2_IdeallySymmetric_InvariantMode}(c) confirms the invariance of the inner-group modes to grid strength changes, while \figref{Fig:Case2_IdeallySymmetric_InvariantMode}(d) validates the invariance of the group-grid mode to group's internal control variations, as the corresponding relative changes remain below 1\%.

\subsection{Case 2: Quasi-Symmetric System}
This case continues the analysis based on the system configuration shown in \figref{Fig:Case2_IdeallySymmetric_Layout}, but with parameters adjusted to reflect more realistic conditions. Specifically, the active power outputs of GFM2 and GFM4 are modified by increasing and decreasing 30\% from their nominal values, respectively. In addition, the line impedances $Z_{2-5}$ and $Z_{4-5}$ are adjusted by 5\% above and below their nominal values, respectively. As a result, the grid-connected system becomes quasi-symmetric.

The pole distribution shown in \figref{Fig:Case23_PoleMap}(b) resembles that in \figref{Fig:Case23_PoleMap}(a), but the inner-group modes are closely located rather than overlapped. The rightmost side LFO modes at 11.8 Hz and 11.3 Hz (both inner-group modes), along with the 3.3 Hz mode (a group-grid mode), are identified as critical modes for participation analysis, as illustrated in \figref{Fig:Case23_PF}(b). In subfigures \#5 and \#6, the participation distribution highlights the inner-group characteristics and exposes the weak robustness problem. This problem is effectively solved using the group participation factors, as presented in subfigure \#8. It is evident from this subfigure that the 11.8 Hz and 11.3 Hz modes are induced by the dynamic interactions between the synchronization control loop and the coupling inductor of inverters, with each inverter contributing in a similar manner. On the other hand, the 3.3 Hz mode corresponds to the group-grid mode, similar to the observation in subfigure \#3.

From \figref{Fig:Case23_PF}(a) and (b), although the group participation factors for the inner-group modes of individual GFMs appear identical or nearly so, this outcome is of engineering significance. It confirms that these modes are collective phenomena arising from the system symmetry, rather than being attributable to any specific apparatus. The group participation factor offers a robust basis for designing group-level damping strategies, as it reveals that mitigating repeated or close modes requires coordinated adjustments across the entire group, rather than targeting individual apparatuses. Hence, the group participation factor delivers essential engineering insight by ensuring a reliable and interpretable group-level participation characterization that traditional participation factor methods cannot provide.

The sensitivities of the conventional and group participation factors are next verified under small parameter perturbations. In \figref{Fig:Case23_PF}(b), Mode 34 exhibits majority participation from GFM2 and minority participation from GFM4, whereas Mode 36 shows the opposite. The line impedances $Z_{2-5}$, $Z_{3-5}$, and $Z_{4-5}$ are perturbed from their original values of $0.019+j0.095$ p.u., $0.02+j0.1$ p.u., and $0.021+j0.105$ p.u. to $0.0198+j0.099$ p.u., $0.0192+j0.096$ p.u., and $0.0206+j0.103$ p.u., respectively. The resulting participation results after perturbation, given in \figref{Fig:Case3_QuasiSymmetric_Perturbation}(a) and (b), indicate that the participation distribution has changed. Mode 34 now has majority participation from GFM3 and minority from GFM4, while Mode 36 has majority participation from GFM4 and minority from GFM3. This discrepancy demonstrates that conventional participation factors are not robust and are highly sensitive to minor perturbations, which are unavoidable in real systems. Such sensitivity poses an engineering problem, as a damping strategy designed based on the initial analysis would be ineffective or even destabilizing after a minor parameter variation. In contrast, the group participation factor validates its superiority by remaining largely unchanged under the same perturbations, as evidenced in \figref{Fig:Case3_QuasiSymmetric_Perturbation}(c). It reliably identifies the entire symmetric group as the target for coordinated control, thereby offering a dependable foundation for system stability enhancement.

In addition, to simplify the analysis and for clarity, the external grid is modeled as an infinite bus with a line impedance. However, the grid structure can be made more complex to investigate a wider range of dynamics. For example, a conventional generation station can be inserted into the external grid to evaluate its impact on low-frequency group-grid modes.

\subsection{Case 3: Group-Symmetric System}
\begin{figure}[t]
\centering
\includegraphics[scale=0.35]{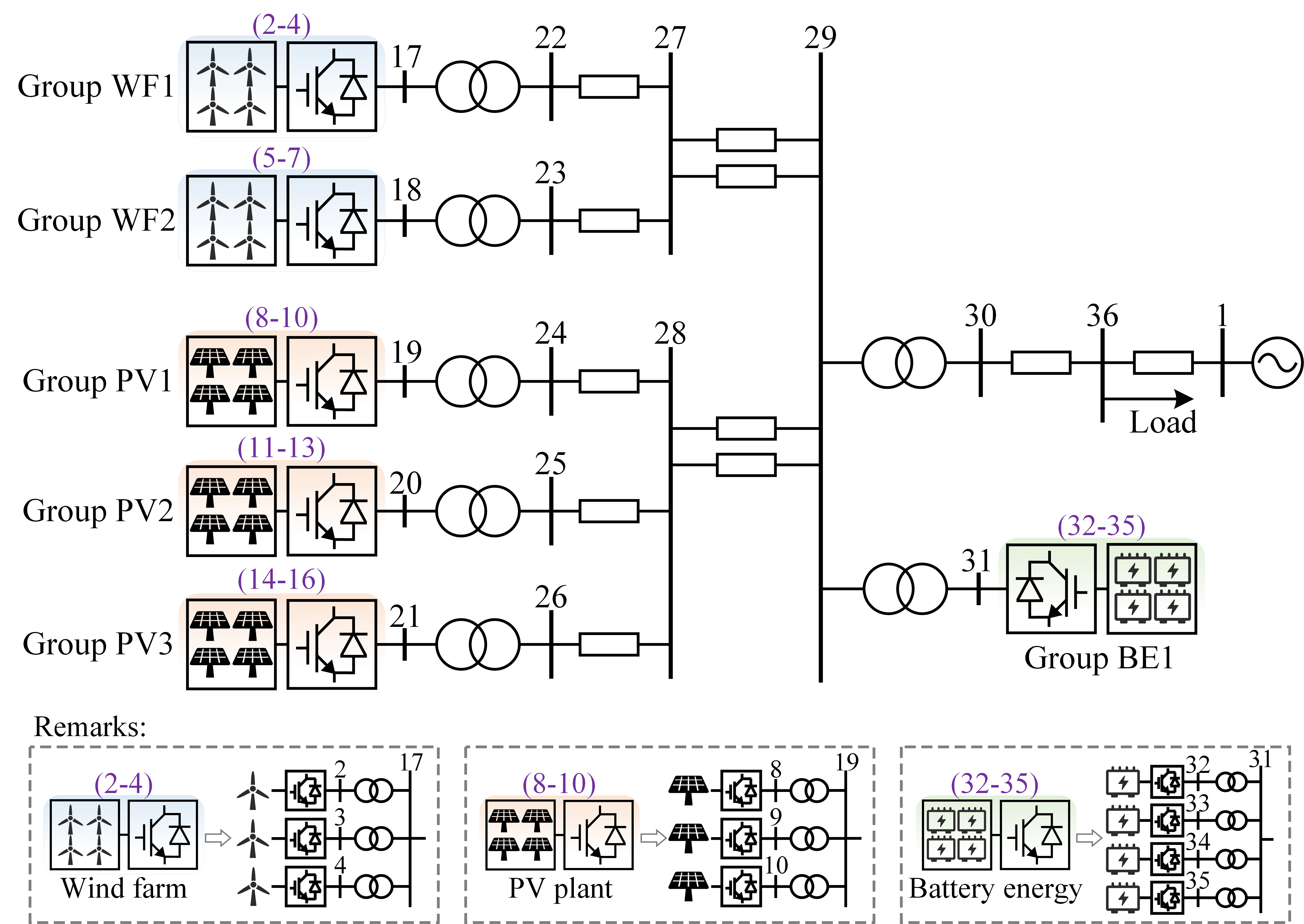}
\caption{Layout of the group-symmetric system.}
\label{Fig:Case4_GroupSymmetric_Layout}
\end{figure}

\begin{figure*}[t]
\centering
\includegraphics[scale=0.44]{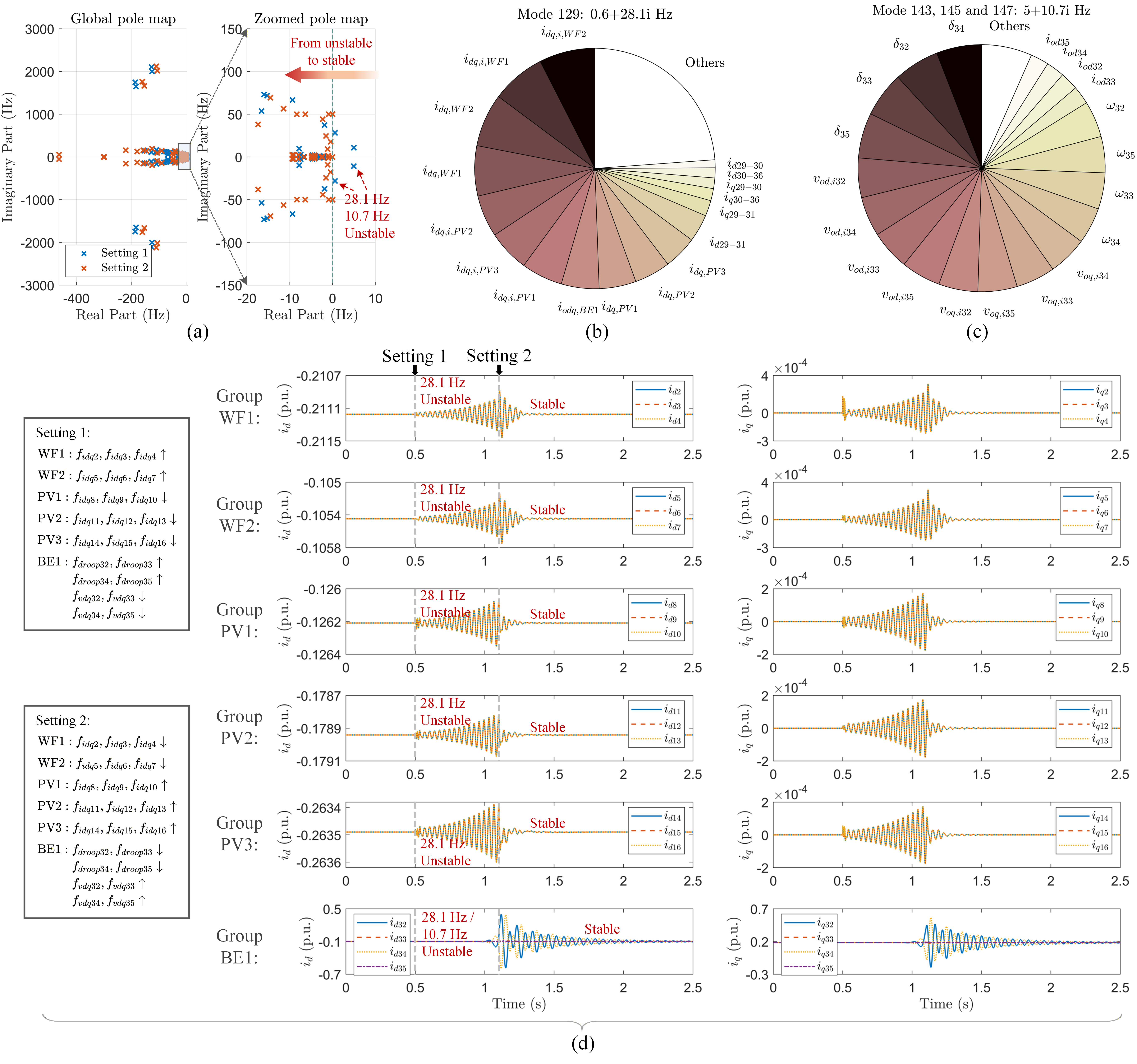}
\caption{Modal analysis results of the group-symmetric system. (a) Pole map. (b) Participation factors for the single unstable mode. (c) Group participation factors for the repeated unstable modes. (d) Simulation results showing the currents of all six groups.}
\label{Fig:Case4_GroupSymmetric_Results}
\end{figure*}
To investigate the modal characteristics of group-symmetric configurations, a modified renewable energy base in Northwest China is selected as a case study\cite{lin2023reactive,liu2025optimal}. The corresponding equivalent topology is shown in \figref{Fig:Case4_GroupSymmetric_Layout}. The system comprises six groups, categorized into three types: wind farm (blue), PV plant (orange), and battery energy (green). Several simplifications are made in the modeling process: Each wind farm or PV plant group contains three generation units, each modeled as a GFL inverter. The battery energy group includes four energy storage units, which are simplified as GFM inverters. Consequently, the modeled system constitutes a mixed GFM-GFL power system.

\figref{Fig:Case4_GroupSymmetric_Results} presents the modal analysis results. As shown in \figref{Fig:Case4_GroupSymmetric_Results}(a), the system under Setting 1 exhibits instability with dominant unstable modes: 28.1 Hz modes identified as group-grid modes, and 10.7 Hz modes that appear repetitively and are categorized as inner-group modes. \figref{Fig:Case4_GroupSymmetric_Results}(b) and (c) show the participation results of these unstable modes: The 28.1 Hz modes are mainly participated by the ac current control loops ($i_{dq}$, $i_{dq,i}$) of all six groups (WF1, WF2, PV1, PV2, PV3, BE1), as well as by the grid currents flowing through Lines 29-30, 29-31, and 30-36. The 10.7 Hz modes are dominated by the synchronization control loop ($\delta,\omega$), the voltage control loop ($v_{odq,i}$), and the coupling inductor ($i_{od}$) of the GFM inverters at Bus 32 to Bus 35 in Group BE1.

Based on the modal insights, Setting 2 is implemented to stabilize the system by fine-tuning the critical control parameters. Specifically, the current control bandwidths ($f_{idq}$) of WF1 and WF2 are reduced, while those of PV1, PV2, and PV3 are increased to mitigate the 28.1 Hz group-grid modes. Meanwhile, the droop control bandwidths ($f_{droop}$) in BE1 are decreased and the voltage control bandwidths ($f_{vdq}$) are increased to suppress the 10.7 Hz inner-group modes. As illustrated in \figref{Fig:Case4_GroupSymmetric_Results}(a), all system poles (highlighted in orange) are located in the left-half of the complex plane, i.e., a stable system.

The modal analysis coincides with the EMT simulations illustrated in \figref{Fig:Case4_GroupSymmetric_Results}(d). Under Setting 1, the 28.1 Hz instability exists in all six groups, while the 10.7 Hz instability only appears in Group BE1. After transitioning to Setting 2, all observed oscillations are effectively damped, and the system ultimately achieves stability.


\section{Conclusion} \label{Section:Conclusion}
This paper investigates the dynamic characteristics introduced by symmetry in renewable energy power systems from the perspective of small-signal stability. Modal analysis, including eigenvalue patterns and participation analysis, is systematically conducted on ideally-, quasi-, and group-symmetric systems. The inner-group modes and group-grid modes are defined. Through the mathematical derivation, it is revealed that the inner-group modes are consistently associated with the repeated modes in ideally-symmetric systems or the close modes in quasi-symmetric systems, whereas the group-grid modes are related to the distinct modes (non-repetitive). The group participation factors are proposed to solve the shortcomings of conventional participation factors, which are failed for repeated and close mode analysis. The invariance properties of inner-group modes indicate that they are most effectively stabilized by adjusting subsystems within the associated group, rather than modifying other parts of the system. The invariance of group-grid modes implies that minor adjustments to a few subsystems within a group have little impact on these modes, as the terminal characteristics seen from the PCC remain largely unchanged. These findings demonstrate that symmetry is not merely a modeling convenience but a system-level property with clear engineering implications. It offers a physically interpretable framework for control design and parameter tuning, which help with stability analysis of power systems with high penetration of IBRs.

The studied renewable energy power system comprises a group (or groups) of subsystems connected in parallel and an external grid. It is noted that a string configuration is more common in practice. The symmetry properties under such circumstance are discussed in \appendixref{Section:Discussion about Symmetry in Practice}.


\appendices

\section{Conventional Participation Factors for Repeated Modes}\label{Section:Conventional Participation Factors for Repeated Modes}
In this paper, it is assumed that any repeated modes in the system are solely caused by symmetry. A key parameter for such modes is the geometric multiplicity, which is defined as the dimension of the eigenspace (or characteristic subspace) corresponding to a given repeated eigenvalue, or equivalently as the number of linearly independent eigenvectors corresponding to the same eigenvalue\cite{strang2016introduction}. The definition is expressed as
\begin{equation}
    g(\lambda_i)=n-\text{rank}(A-\lambda_iI),
\end{equation}
where $A$ is the system state matrix, $\lambda_i$ is the $i$-th eigenvalue, $n$ is the dimension of $A$, and $I$ represents the identity matrix. When $g(\lambda_i)>1$, the eigenvalue $\lambda_i$ is associated with multiple linearly independent eigenvectors, thereby forming a multi-dimensional characteristic subspace. In this case, the participation factor for the repeated mode is not unique. We can obtain $g(\lambda_i)$ different participation factors, depending on the specific choice of eigenvectors within the characteristic subspace.

To illustrate this concept, consider the following example.
\begin{equation}
    \label{Appendix_Example_A1}
    A=\begin{bmatrix}
        2 & 0 \\
        0 & 2
    \end{bmatrix}
\end{equation}
The repeated modes are $\lambda_1=\lambda_2=2$. Taking $\lambda_1$ as an example, its geometric multiplicity is $g(\lambda_1)=2>1$. Hence, the eigenvalue $\lambda_1$ is associated with two linearly independent right eigenvectors. Any linear combination of the basis vectors
\begin{equation}
    v_1=\begin{bmatrix}
        1\\
        0
    \end{bmatrix}
    \text{and}\;
    v_2=\begin{bmatrix}
        0\\
        1
    \end{bmatrix}
\end{equation}
can serve as right eigenvectors. If the right eigenvectors are chosen as $\phi_1=\left[1\;0\right]^\mathrm{T}$ and $\phi^{\prime}_1=\left[0\;1\right]^\mathrm{T}$, and the left eigenvectors are chosen as $\psi_1=\left[1\;0\right]$ and $\psi^{\prime}_1=\left[0\;1\right]$, the resulting participation factor for $\lambda_1$ is not unique.

For the state matrix of the simple example in \eqref{RLStateMatrix}, the geometric multiplicity of the repeated modes is $2$. For the state matrix of the general derivation in \eqref{StateMatrix}, the geometric multiplicity is $M-1$. In both cases, the conventional participation factors for the repeated inner-group modes are ill-defined.

Another key parameter for the repeated modes is the algebraic multiplicity, defined as the multiplicity of a corresponding eigenvalue as a root of the characteristic polynomial. Therefore, for the repeated modes in both the simple example and the general derivation of ideally-symmetric systems, the algebraic multiplicity equals the geometric multiplicity.


\section{Group Participation Factors for Repeated Modes}\label{Section:Group Participation Factors for Repeated Modes}
Suppose that the state matrix $A$ has $n_g$ multiple eigenvalues $\lambda_i,i=1,2,\cdots,n_g$. The group participation factor of the $k$-th state variable to the repeated modes is $gpf_{k}=\sum_{i=1}^{n_g}\Psi_{ik}\Phi_{ki}$. If a new set of right eigenvectors (column vectors with the superscript ``$^{\prime\prime}$'') is selected, they share the same characteristic subspace as the original set. Consequently, there exists an invertible matrix $Q\in \mathbb{C}^{n_g\times n_g}$ such that
\begin{equation}
    \left[ \phi^{\prime\prime}_1 \; \phi^{\prime\prime}_2 \; \cdots \; \phi^{\prime\prime}_{n_g}\right]=\left[ \phi_1 \; \phi_2 \; \cdots \; \phi_{n_g}\right]Q.
\end{equation}
Accordingly, the individual transformed right eigenvectors and their components can be expressed as
\begin{equation}
    \phi_{i}^{\prime\prime}=\sum_{j=1}^{n_g} \phi_{j} Q_{ji},\;\Phi_{ki}^{\prime\prime}=\sum_{j=1}^{n_g} \Phi_{kj} Q_{ji}.
\end{equation}
Based on the normalized conditions $\psi_{i}\phi_{i}=\psi_{i}^{\prime\prime}\phi_{i}^{\prime\prime}=1$, the left eigenvectors (row vectors) and their components satisfy
\begin{equation}
    \psi_{i}^{\prime\prime}=\sum_{j=1}^{n_g} Q^{-1}_{ij} \psi_{j},\;\Psi_{ik}^{\prime\prime}=\sum_{j=1}^{n_g} Q^{-1}_{ij} \Psi_{jk}.
\end{equation}
With the new left and right eigenvectors, the group participation factor is reformulated as
\begin{equation}
    \begin{aligned}
        gpf^{\prime\prime}_{k}&=\sum_{i=1}^{n_g}\Psi^{\prime\prime}_{ik}\Phi^{\prime\prime}_{ki}
        =\sum_{i=1}^{n_g}\sum_{j=1}^{n_g} Q^{-1}_{ij} \Psi_{jk}\sum_{l=1}^{n_g} \Phi_{kl} Q_{li}\\
        &=\sum_{j=1}^{n_g}\sum_{l=1}^{n_g}\Psi_{jk}\Phi_{kl}\sum_{i=1}^{n_g}Q_{li}Q^{-1}_{ij}
        =\sum_{j=1}^{n_g}\Psi_{jk}\Phi_{kj}=gpf_k.
    \end{aligned}
\end{equation}
Therefore, although the participation factors for repeated modes may vary depending on the specific choice of eigenvectors, the group participation factors remain invariant and independent of the eigenvector selection.

\section{Group Participation Factors for Close Modes}\label{Section:Group Participation Factors for Close Modes}
Before investigating the group participation factors for close modes, we first introduce the mathematical concept of the Riesz spectral projector, defined via a spectral integral\cite{kato1995perturbation}. For an isolated eigenvalue $\lambda_i$ of a matrix $A$, the associated spectral projector is given by
\begin{equation}
    P_{i}=\frac{1}{2\pi i}\oint_{\Gamma_i}(sI-A)^{-1}\,ds,
\end{equation}
where $\Gamma_i$ is a positively oriented contour enclosing $\lambda_i$ but no other eigenvalues of $A$. When $\lambda_i$ is simple (i.e., its algebraic multiplicity is $1$), this projector reduces to $ P_{i}=\phi_{i}\psi_{i}$, with $\phi_{i}$ and $\psi_{i}$ denoting the right and left eigenvectors, respectively\cite{dunford1958linear}. Recalling the definition of the participation factor $pf_{ki}=\Psi_{ik}\Phi_{ki}$, it follows that
\begin{equation}
    pf_{ki}=(P_{i})_{kk}.
\end{equation}
This indicates that the participation factor $pf_{ki}$ is the $k$-th diagonal element of the spectral projector $P_{i}$.

For a set of close modes $\mathcal{C}=\{\lambda_1,\cdots,\lambda_{n_g}\}$, the total spectral projector is defined as
\begin{equation}
    P_{\mathcal{C}}=\frac{1}{2\pi i}\oint_{\Gamma}(sI-A)^{-1}\,ds,
\end{equation}
where $\Gamma$ is a contour that encloses all eigenvalues in $\mathcal{C}$ while excluding the remainder of the spectrum. By applying the residue theorem to the resolvent $(sI-A)^{-1}$, the total spectral projector satisfies $P_{\mathcal{C}}=\sum_{\lambda_i \in \mathcal{C}}P_{i}$.
Given the definition of the group participation factor $gpf_{k}=\sum_{i=1}^{n_g}\Psi_{ik}\Phi_{ki}$, we obtain
\begin{equation}
    gpf_{k}=(P_{\mathcal{C}})_{kk}.
\end{equation}
This means that the group participation factor $gpf_{k}$ is the $k$-th diagonal element of the total spectral projector $P_{\mathcal{C}}$.

Therefore, if we wish to prove that the group participation factor $gpf_{k}$ is robust under parameter variations, we can shift to proving that the total projection operator $P_{\mathcal{C}}$ is robust, which has already been proved in perturbation theory and will not be repeated here\cite{kato1995perturbation, dunford1958linear}.

\begin{figure*}[t]
\centering
\includegraphics[scale=0.64]{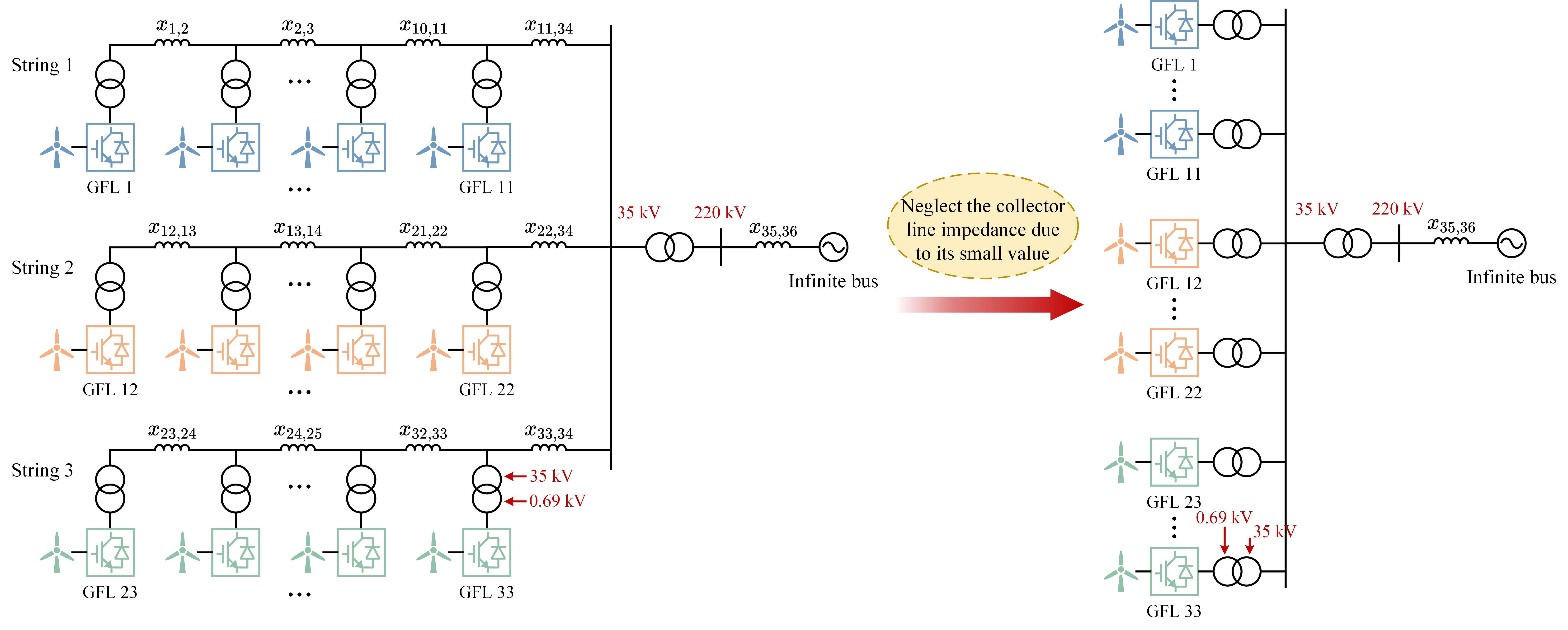}
\caption{Network topology of a renewable energy power system, using a wind farm as an example. The left-hand side is a string configuration (in practice) while the right-hand side is a parallel configuration (simplified analysis).}
\label{Fig:WindFarmTopology}
\end{figure*}

\begin{figure}[t]
\centering
\includegraphics[scale=0.5]{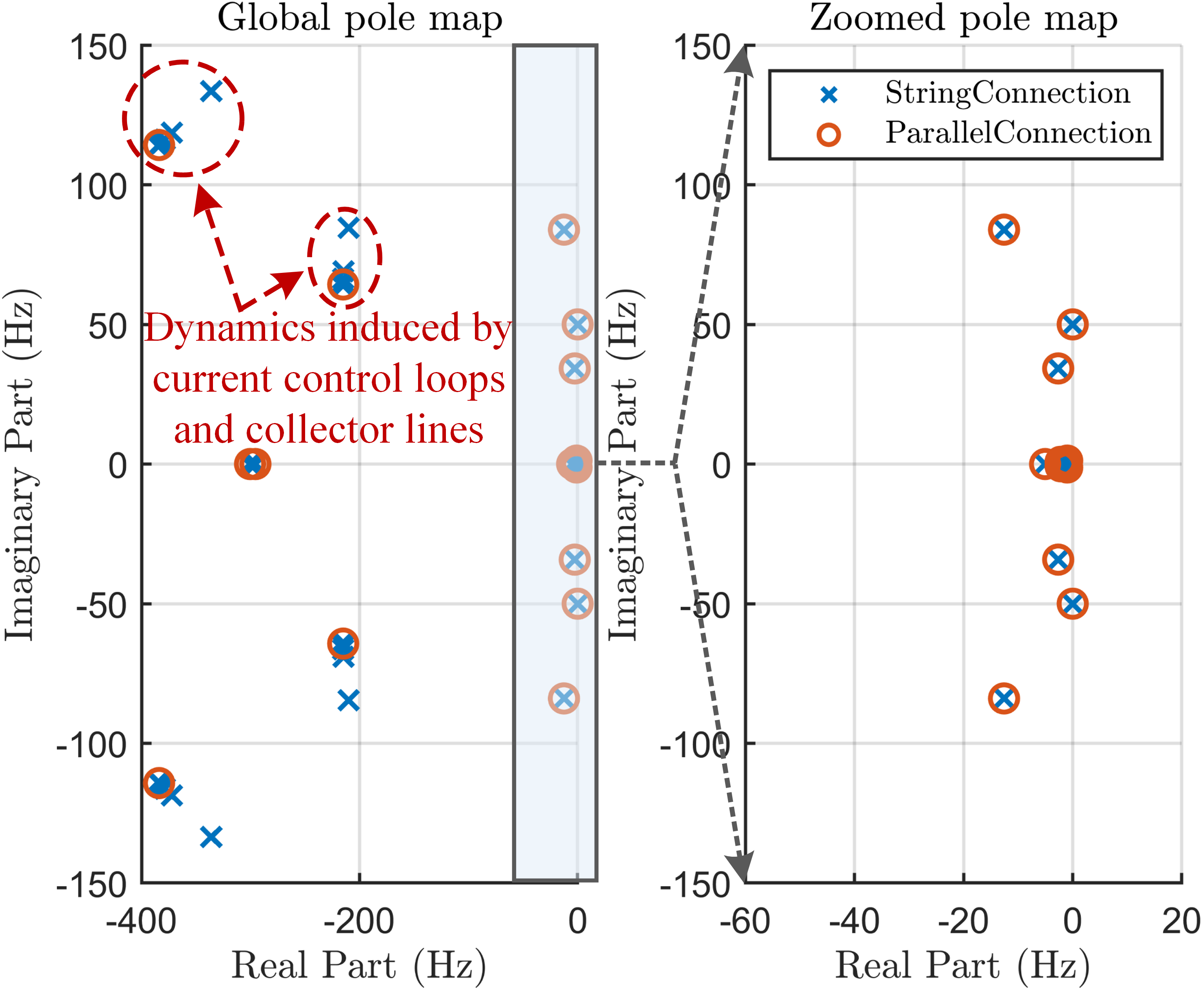}
\caption{Comparison of pole maps of the wind farm system under string and parallel configurations.}
\label{Fig:WindFarmPoleMap}
\end{figure}

\section{Discussion about Symmetry in Practice}\label{Section:Discussion about Symmetry in Practice}
This paper investigates the eigenvalue patterns and participation factors in renewable energy power systems based on symmetry, which mainly denotes that the subsystems within the parallel connection segment possess identical or nearly identical structure and parameters. In practical, generation units are often homogeneous, leading to such symmetric conditions. However, they are connected in a series string, as illustrated in \figref{Fig:WindFarmTopology}\cite{zhou2023small}. In this configuration, multiple generation units are connected via a collector line, whose impedance is usually small in practice. If we neglect the collector network impedance, the system becomes equivalent to a parallel connection. This equivalent modeling eliminates the requirement to aggregate all turbines on a string. For simplicity, each generation unit in the wind farm is modeled in the subsequent analysis as a GFL inverter with a step-up transformer.

\figref{Fig:WindFarmPoleMap} compares the closed-loop poles of the entire system for the string and parallel configurations. The results show that the poles of the two configurations are almost overlapping, with minor differences in the mid-to-high frequency range. These differences are induced by the current control loops and collector line impedances; they become smaller as the collector line impedances decrease. Therefore, the parallel connection, which is the focus of this paper, can highlight the essential symmetric properties and simplify the analysis while remaining consistent with the dynamics of the more practical string configuration, especially for synchronization-related dynamics. A detailed investigation of these minor differences, including a comprehensive comparison of the two configurations, is reserved for future work.


\ifCLASSOPTIONcaptionsoff
  \newpage
\fi

\bibliographystyle{IEEEtran}
\bibliography{Paper}

\end{document}